\documentclass[12pt,iop,numberedappendix]{emulateapj}
\usepackage{epsfig,graphicx}

\newcommand{\etal}{et al.}

\newcommand{\hone}{\hbox{{\rm H{\small I}}}}
\newcommand{\htwo}{\hbox{${\rm H_2}$}}
\newcommand{\cone}{\hbox{${\rm CO \ {\it J}=1-0}$}}
\newcommand{\ctwo}{\hbox{${\rm CO \ {\it J}=2-1}$}}

\newcommand{\nmol}{\hbox{$\rm N_{mol}$}}
\newcommand{\xco}{\hbox{$\rm X_{CO}$}}
\newcommand{\um}{\hbox{$\mu$m}}
\newcommand{\Sstar}{\hbox{$\Sigma_{\ast}$}}
\newcommand{\Shone}{\hbox{$\Sigma_{\rm HI}$}}
\newcommand{\Shtwo}{\hbox{$\Sigma_{\rm H2}$}}
\newcommand{\Sgas}{\hbox{$\Sigma_{\rm gas}$}}

\newcommand{\Ssfr}{\hbox{$\Sigma_{\rm SFR}$}}

\newcommand{\Smir}{\hbox{$\Sigma_{24\,\um}$}}

\newcommand{\kmsec}{\mbox{km \ sec$^{-1}$}}

\newcommand{\sigint}{\mbox{$\rm \sigma_{int}$}}
\newcommand{\sigmes}{\mbox{$\rm \sigma_{mes}$}}
\newcommand{\sigobs}{\mbox{$\rm \sigma_{obs}$}}

\newcommand{\msun}{\hbox{$\rm M_{\odot}$}}
\newcommand{\mstar}{\hbox{$\rm M_{\ast}$}}

\newcommand{\lstarks}{\hbox{$\rm L_{K_s}$}}
\newcommand{\lsunks}{\hbox{$\rm L_{\odot, K_s}$}}
\newcommand{\mlrks}{\hbox{$\Upsilon_{\ast}^{K_s}$}} 
\newcommand{\mhtwo}{\hbox{$\rm M_{H_2}$}}

\newcommand{\msunpc}{\hbox{$\rm M_{\odot} \ pc^{-2}$}} 
\newcommand{\msunyr}{\mbox{$\rm M_{\odot} \ yr^{-1}$}}
\newcommand{\msunyrpc}{\mbox{$\rm M_{\odot} \ Gyr^{-1} \ pc^{-2}$}}
\newcommand{\ergs}{\mbox{\rm erg~s$^{-1}$}}

\newcommand{\tdep}{\hbox{$\tau_{\rm dep}^{\rm mol}$}}
\newcommand{\avetdep}{\hbox{$\rm {\bar{\tau}}_{dep}^{mol}$}}

\newcommand{\oabundance}{\hbox{$\rm 12+\log(O/H)$}}

\newcommand{\resolution}{\hbox{$\mathcal R$}}
\newcommand{\thubble}{\hbox{$\rm \tau_{Hubble}$}}
\newcommand{\tjeans}{\hbox{$\rm \tau_J^{mol}$}}
\newcommand{\tff}{\hbox{$\rm \tau_{ff}^{mol}$}}
\newcommand{\tsa}{\hbox{$\rm \tau_\ast$}}
\newcommand{\hzstar}{\hbox{$\rm h_{z,\ast}$}}
\newcommand{\hrstar}{\hbox{$\rm h_{R,\ast}$}}
\newcommand{\cstar}{\hbox{$\rm c_\ast$}}
\newcommand{\cgas}{\hbox{$\rm c_g$}}
\newcommand{\ssfr}{\hbox{$\rm sSFR$}}

\shorttitle{Molecular Gas Star Formation Law in the CARMA STING Survey}
\shortauthors{Rahman et al.}

\begin{document}
\title{CARMA Survey Toward Infrared-bright Nearby Galaxies (STING) 
II: Molecular Gas Star Formation Law and Depletion Time Across the
Blue Sequence}

\author{
Nurur Rahman\altaffilmark{1}, 
Alberto D. Bolatto\altaffilmark{1},
Rui Xue\altaffilmark{2},
Tony Wong\altaffilmark{2}, 
Adam K. Leroy\altaffilmark{3}, 
Fabian Walter\altaffilmark{4},
Frank Bigiel\altaffilmark{5}, 
Erik Rosolowsky\altaffilmark{6}, 
David B. Fisher\altaffilmark{1},
Stuart N. Vogel\altaffilmark{1},
Leo Blitz\altaffilmark{7}, 
Andrew A. West\altaffilmark{8}, and
J\"{u}rgen Ott\altaffilmark{9}
}
\altaffiltext{1}{Department of Astronomy, University of Maryland, 
College Park, MD, USA; nurur@astro.umd.edu}
\altaffiltext{2}{Department of Astronomy, University of Illinois, 
Urbana-Champaign, IL, USA}
\altaffiltext{3}{National Radio Astronomy Observatory, Charlottesville, 
VA, USA}
\altaffiltext{4}{Max-Planck-Institute fur Astronomie, Konigstuhl 17, 
Heidelberg, Germany}
\altaffiltext{5}{Institut f\"ur Theoretische Astrophysik, 
Universit\"at Heidelberg, Albert-Ueberle Str. 2, 69120 Heidelberg, 
Germany}
\altaffiltext{6}{I. K. Barber School of the Arts \& Science, University 
of British-Columbia, Kelowna, BC, Canada}
\altaffiltext{7}{Department of Astronomy, University of California, 
Berkeley, CA, USA}
\altaffiltext{8}{Department of Astronomy, Boston University, Boston, MA, 
USA}
\altaffiltext{9}{National Radio Astronomy Observatory, Socorro, NM, USA}
\begin{abstract}
We present an analysis of the relationship between molecular gas and
current star formation rate surface density at sub-kpc and kpc scales
in a sample of 14 nearby star-forming galaxies. Measuring the
relationship in the bright, high molecular gas surface density
($\Shtwo\gtrsim$20~\msunpc) regions of the disks to minimize the
contribution from diffuse extended emission, we find an approximately
linear relation between molecular gas and star formation rate surface
density, $\nmol\sim0.96\pm0.16$, with a molecular gas depletion time,
$\tdep\sim2.30\pm1.32$~Gyr. We show that, in the molecular regions of
our galaxies there are no clear correlations between \tdep\ and the
free-fall and effective Jeans dynamical times throughout the sample.
We do not find strong trends in the power-law index of the spatially
resolved molecular gas star formation law or the molecular gas
depletion time across the range of galactic stellar masses sampled
(\mstar~$\sim$$10^{9.7}-10^{11.5}$~\msun). There is a trend, however,
in global measurements that is particularly marked for low mass
galaxies. We suggest this trend is probably due to the low surface
brightness \cone, and it is likely associated with changes in
CO-to-H$_2$ conversion factor.
\end{abstract}

\section{Introduction}
It is well known that the star formation activity in a galaxy
correlates strongly with its stellar mass or stellar surface density
(Dopita \& Ryder 1994; Hunter \etal\ 1998; Kauffmann \etal\ 2003;
Blitz \& Rosolowsky 2004, 2006), and only weakly on galaxy morphology
(Boselli \etal\ 2001). Characterizing the relations between stellar
mass, star formation rate (SFR), and gas densities over cosmic time 
will provide important constraints on galaxy evolution by connecting 
the past history, present activity, and future growth of a galaxy 
(Schiminovich \etal\ 2010). 

A key factor in galaxy evolution is the rate at which molecular gas 
is converted to stars. Observational studies find that the relationship 
between SFR and gas content in galaxies can be written in the form, 
$\rm \Sigma_{\rm SFR} = A \ \Sigma_{\rm gas}^N$, where \Ssfr\ and 
\Sgas\ are the SFR and gas surface densities 
respectively, $\rm A$ is the normalization constant representing the
efficiency of the process, and $\rm N$ is the power-law index (Schmidt
1959; Kennicutt 1989, 1998). The gas can be atomic (\hone) or
molecular (\htwo) or a combination of both (\hone+\htwo). This
relationship is generally known as the Kennicutt-Schmidt law or the
star formation law, and variations on it are used as a empirical
recipe in galaxy modeling (Schaye \& Della Vecchia 2008; Lagos \etal\
2010).

In recent years has become increasingly clear the importance of
understanding the link between star formation activity and
gravitationally bound molecular clouds (Wong \& Blitz 2002; Tutukov 
2006; Kennicutt \etal\ 2007; Bigiel \etal\ 2008; Leroy \etal\ 2008, 
hereafter L08; Blanc \etal\ 2009; Verley \etal\ 2010; Onedara \etal\ 
2010; Bigiel \etal\ 2011; Schruba \etal\ 2011). 
Recent observational studies indicate 
a strong relationship between the \hone-to-\htwo\ transition and
the gravitational potential of the stellar disk and thus evince the 
connection between stellar pressure and the phase transition in the 
interstellar medium (Wong \& Blitz 2002; Blitz \& Rosolowsky 2004, 
2006). Numerical simulations that include cold gas and heating from 
young stellar population in evolution of interstellar medium (ISM) 
are in general agreement with the observations (Robertson \& Kravtsov 
2008).

Most properties of galaxies, not least their star formation activity,
are strongly correlated with mass (see Gavazzi 2009 for a
review). Because new stars are born inside giant molecular clouds
(GMCs), which themselves evolve within the existing galactic stellar
potential (Rafikov 2001; Li \etal\ 2005, 2006), it is important to
characterize the interconnections between stellar mass, \htwo, and SFR
(see Shi \etal\ 2011 for a recent study). To this end, we investigate
the \Shtwo-\Ssfr\ relation at sub-kpc and kpc scales in 14 nearby
star-forming disk galaxies. While gas-SFR surface density relation
provides an understanding of on-going activities in the disk, the
molecular gas depletion time (\tdep) provides a measure of its future
evolution. It is a quantitative measure of the efficiency of the star
formation activity in molecular clouds, defined as the time required
for available molecular gas to be converted into stars while
maintaining the existing rate of star formation (Roberts 1963; Larson
\etal\ 1978; Kennicutt 1983; Kennicutt \etal\ 1994).

\tabletypesize{\scriptsize} 
\begin{deluxetable*}{@{}l@{}c@{}c@{}c@{}c@{}c@{}c@{}c@{}r@{$\pm$}lr@{$\pm$}lr@{$\pm$}lr@{$\pm$}lr@{$\pm$}l@{}@{}c@{}l@{}}
\tablecaption{Basic Information for the STING Sample}
\tablewidth{7.0truein}
\tablecolumns{20}
\tablehead{
\colhead{Object}   
&\colhead{Hubble} 
&\colhead{$\rm D$} 
&\colhead{$\rm D_{25}$} 
&\colhead{\resolution}
&\colhead{$\mathcal S$}
&\colhead{${\mstar}$} 
&\colhead{${\mhtwo}$} 
&\multicolumn{2}{c}{\avetdep}    
&\multicolumn{2}{c}{$\rm \log A$}    
&\multicolumn{2}{c}{\nmol}    
&\multicolumn{2}{c}{$\rm \log A$}    
&\multicolumn{2}{c}{\nmol}    
&\colhead{Corr.}
&\colhead{Corr.} \\
\colhead{}
&\colhead{Type} 
&\colhead{Mpc}  
&\colhead{kpc}
&\colhead{pc}
&\colhead{kpc}
&\colhead{\msun}
&\colhead{\msun}
&\multicolumn{2}{c}{Gyr}    
&\multicolumn{2}{c}{6\arcsec}    
&\multicolumn{2}{c}{6\arcsec}    
&\multicolumn{2}{c}{1 kpc}    
&\multicolumn{2}{c}{1 kpc}   
&\colhead{$\rho$}
&\colhead{$\rho$} \\
\colhead{(1)}
&\colhead{(2)} 
&\colhead{(3)}
&\colhead{(4)}  
&\colhead{(5)}  
&\colhead{(6)}
&\colhead{(7)}
&\colhead{(8)}
&\multicolumn{2}{c}{(9)}    
&\multicolumn{2}{c}{(10)}    
&\multicolumn{2}{c}{(11)}    
&\multicolumn{2}{c}{(12)}    
&\multicolumn{2}{c}{(13)}   
&\colhead{(14)}
&\colhead{(15)}
} 
\startdata 
 NGC~628   &~~~SA(s)c      &~7.30$^{\dag}$   &21.23  &~211       &~4.24  &~9.95  &7.81  &2.91 &0.75                   &-0.23  &0.05                   &0.86  &0.04        &\multicolumn{2}{c}{\ldots}
  &\multicolumn{2}{c}{\ldots}  &+0.31 &-0.24        \\
 NGC~772   &~~~SA(s)b      &32.00            &51.77  &{\bf~925}  &18.62  &11.39  &9.71  &7.20 &1.59\tablenotemark{a}  &-0.59  &0.19\tablenotemark{a}  &0.88  &0.10$^{a}$  &-0.59  &0.19   &0.88  &0.10       &-0.40 &-0.67\\
 NGC~1637  &~SAB(rs)c      &10.92            &11.38  &~316       &~6.35  &~9.83  &8.25  &2.62 &0.84                   & 0.07  &0.11\tablenotemark{b}  &0.69  &0.08$^{b}$  &-0.20  &0.22   &1.16  &0.13$^{c}$ &-0.54 &-0.66\\
 NGC~3147  &~SA(rs)bc      &43.13            &43.39  &{\bf1247}  &25.09  &11.36  &9.90  &6.16 &1.35\tablenotemark{a}  &-0.51  &0.10\tablenotemark{a}  &0.87  &0.05$^{a}$  &-0.51  &0.10   &0.87  &0.05       &-0.25 &-0.62\\
 NGC~3198  &~SB(rs)c       &13.70$^{\dag}$   &21.11  &~396       &~7.97  &10.25  &8.14  &1.18 &0.28                   &-0.34  &0.16                   &1.16  &0.09        & 0.28  &0.03   &0.78  &0.02       &+0.25 &+0.09\\
 NGC~3593  &~SA0/a?(s)     &~5.50            &~5.12  &~159       &~3.20  &~9.81  &8.21  &1.98 &1.00                   &-0.60  &0.09                   &1.20  &0.04        &-0.65  &0.03   &1.23  &0.02       &+0.37 &+0.14\\
 NGC~3949  &~SA(s)bc?      &18.41            &11.74  &~532       &10.71  &10.00  &8.22  &5.16 &1.61                   &\multicolumn{2}{c}{\ldots}     &\multicolumn{2}{c}{\ldots} &\multicolumn{2}{c}{\ldots} 
 &\multicolumn{2}{c}{\ldots} &+0.34 &-0.25\\
 NGC~4254  &~~~SA(s)c      &16.60$^{\dag}$   &24.33  &~480       &~9.66  &10.48  &9.44  &3.71 &0.94\tablenotemark{a}  &-0.24  &0.06\tablenotemark{a}  &0.86  &0.03$^{a}$  &-0.20  &0.11   &0.82  &0.05       &-0.50 &-0.61\\
 NGC~4273  &~~~SB(s)c      &33.68            &18.50  &{\bf~974}  &19.59  &10.35  &9.34  &1.41 &0.36                   &-0.01  &0.06                   &0.93  &0.03        &-0.01  &0.06   &0.93  &0.03       &-0.24 &-0.34\\
 NGC~4536  &SAB(rs)bc      &14.40$^{\dag}$   &20.66  &~416       &~8.38  &10.24  &8.95  &1.13 &0.37                   & 0.04  &0.10                   &0.96  &0.05        &-0.03  &0.11   &0.98  &0.06       &-0.20 &-0.25\\
 NGC~4654  &SAB(rs)cd      &14.10$^{\ddag}$  &17.55  &~472       &~8.20  &10.21  &8.77  &2.11 &0.54                   &-0.19  &0.05                   &0.93  &0.03        &-0.18  &0.07   &0.84  &0.04       &-0.16 &-0.37\\
 NGC~5371  &SAB(rs)bc      &35.25            &39.89  &{\bf1019}  &20.51  &11.30  &8.94  &4.34 &1.16                   &-0.25  &0.09                   &0.75  &0.06        &-0.25  &0.09   &0.75  &0.06       &-0.61 &-0.65\\
 NGC~5713  &SAB(rs)bc pec  &29.40$^{\dag}$   &22.63  &~850       &17.10  &10.59  &9.59  &1.67 &0.85                   &-0.32  &0.06                   &1.09  &0.03        &-0.29  &0.11   &1.05  &0.05       &+0.15 &~0.00\\
 NGC~6951  &SAB(rs)bc      &22.83            &23.53  &~660       &13.28  &10.77  &9.37  &3.04 &1.33                   &-0.43  &0.06                   &1.00  &0.03        &-0.50  &0.07   &1.03  &0.03       &+0.10 &-0.17
\enddata 
\tablecomments{
Column $(1)$: Galaxy name. 
Column $(2)$: Galaxy Morphology from the RC3 catalog (de Vaucouleurs \etal\ 1991). 
Column $(3)$: Distance ($\rm D$) in Mpc ($^{\dag}$Prescott \etal\ 2007; $^{\ddag}$Tully \etal\ 2009). 
Column $(4)$: Optical diameter ($\rm D_{25}$) in kpc.
Column $(5)$: Spatial resolution (\resolution) in parsec corresponding to 6\arcsec\ angular 
resolution at the adopted distance.
Column $(6)$: Physical extent ($\mathcal S$) of the central arc-minute (in diameter) of the 
disk in kpc. 
Column $(7)$: Logarithmic stellar mass (\msun).
Column $(8)$: Logarithmic molecular gas mass in \msun\ estimated within $\mathcal S$ including 
the contribution of helium. 
Column $(9)$: Average molecular gas depletion time (\avetdep) and its $1\sigma$ dispersion in 
Gyr at 6\arcsec\ resolution. 
Column $(10)$ \& $(11)$: Normalization constant and power-law index (\nmol) of molecular gas 
SF law at 6\arcsec\ resolution.
Column $(12)$ \& $(13)$: Normalization constant and power-law index at 1 kpc resolution. 
Column $(14)$: Spearman's rank correlation coefficient ($\rho$) from \tdep~-~\tjeans\ relation 
at 6\arcsec\ resolution. 
Column $(15)$: Spearman's rank correlation coefficient ($\rho$) from \tdep-\tff\ relation at 
6\arcsec\ resolution. \\ 
The 6\arcsec\ resolution of galaxies NGC~772, NGC~3147, NGC~4273, and NGC~5371, shown in 
bold face, corresponds to ${\cal R}\gtrsim1$~kpc in spatial scale. \\
Stellar mass (\mstar ) in Column (7) is derived from Two Micron All Sky Survey (2MASS) 
$\rm K_s$-band total magnitude with a range of mass-to-light ratio, 
$\mlrks \sim (0.3-0.8)~\msun/\lsunks$, varying from galaxy to galaxy. It is this \mstar\ 
what we call as the integrated stellar mass which incorporate the flux from the entire 
optical disk. The uncertainty is few percent in 
\mstar\ and \mhtwo.\\
$^{a}$ Parameters are derived for the surface density limit, $\Shtwo\gtrsim70$~\msunpc.\\
$^{b}$ Parameters are derived from \Shtwo\ and \Smir\ maps with masked central region. The 
corresponding parameters for the unmasked central region are $\log A$~$\sim$(-1.20$\pm$0.09) 
and \nmol~$\sim$(1.62$\pm$0.05).\\
$^{c}$ Central region was not masked at 1 kpc resolution.
\label{basic_table}}
\end{deluxetable*}

In this study we measure \tdep\ in a sample of galaxies from the Survey 
Toward Infrared-bright Nearby Galaxies (STING), observed in CO with the 
Combined Array for Research in Millimeter-wave Astronomy (CARMA), to 
investigate its relation with stellar mass and various dynamical 
timescales associated with gravitational instability. 
The STING sample is composed of 23 northern ($\delta > -20^{\circ}$),
moderately inclined ($i <75^{\circ}$), high metallicity
(\oabundance~$> 8.1$) galaxies within 45 Mpc. These blue-sequence
star-forming galaxies (Salim \etal\ 2007) have been selected to have
uniform coverage in stellar mass ($10^{8.1}<\mstar/\msun<10^{11.5}$),
star formation activities, and morphological types (A. D. Bolatto
\etal, 2011, in preparation). This is the second in a series of papers
dedicated to exploiting the CO STING data set. In a previous study we
investigated the impact of methodology on the determination of the
spatially resolved molecular gas star formation law in NGC~4254, a
member of the STING survey (Rahman \etal\ 2011; hereafter Paper I). In
this study we explore relationships among current SFR, molecular gas,
and stellar mass in normal star-forming galaxies using similar data
analysis methodologies as in Paper I.

The organization of the paper is as follows. In $\S$\ref{data} 
we briefly present multi-wavelength data set. In $\S$\ref{analysis} we 
provide a description of the data products and data analysis methodology. 
Our main results and a general discussion are given $\S$\ref{results} 
and $\S$\ref{discussion}, respectively. The conclusions are given in 
$\S$\ref{summary}.

\section{Data}
\label{data}
The \cone\ maps are obtained as part of the STING survey using the CARMA 
interferometer. While the full description of the CARMA observations will 
be presented in a forthcoming paper (A. D. Bolatto \etal, 2011, in 
preparation), we briefly mention here some aspects of the interferometric 
data. The maps have a full sensitivity field-of-view of~$\sim$2\arcmin\ 
diameter, and are performed using a 19-pointing hexagonal mosaicing pattern 
with~$\sim$26\arcsec\ spacing.  The angular resolution, measured as the full 
width at half maximum (FWHM) of the synthesized beam of the interferometer, 
varies from galaxy to galaxy in the range $\theta\sim$3\arcsec-5\arcsec. To
construct molecular gas column density maps, we use the \cone\ spectral cube 
to produce integrated \cone\ intensity maps. To avoid integrating over noisy 
channels we use a velocity masking technique similar to that used by the 
BIMA Survey of Nearby Galaxies (SONG; Helfer \etal\ 2003).

To construct SFR tracer maps we use mid-infrared (MIR) 24~\um\ images from 
the Multi-band Imaging Photometers (MIPS; Rieke \etal\ 2004) instrument 
on board the {\em Spitzer Space Telescope}. The calibrated images are 
obtained from the {\em Spitzer} Heritage Archive. 
We further process them by masking bright point sources such as 
Active Galactic Nuclei (AGN) which sometimes dominate the emission in the 
galaxy centers. The images of NGC~1569, NGC~1637, NGC~3147, NGC~3949, 
NGC~5371, and NGC~6503 contain either foreground or background point 
sources, which we mask after visual inspection. The central regions of 
NGC~1637 and NGC~3198 contain bright sharply peaked point-like sources, 
despite the fact that galaxies are not known as AGN in the literature. 
The shapes of these central sources are consistent with the point spread 
function of the MIPS instrument, which has a FWHM of $\sim6\arcsec$. To 
remove the contributions of the point sources we mask the central regions 
of these two galaxies. Four STING galaxies (NGC~3147, NGC~3486, NGC~4151,
and NGC~5371) harbor AGN at their centers, which we also masked.

The maps of stellar mass are constructed from Two Micron All Sky
Survey (2MASS; Skrutskie \etal\ 2006) NIR 2.2~\um ($\rm K_s$-band)
images. The image resolution is 2\arcsec\ for $\rm K_s$-band images. At 
this wavelength masking of foreground and background objects was necessary
for several STING sources.
\section{Data Analysis}
\label{analysis} 
We select 14 galaxies from the 23 star-forming galaxies in the 
STING sample. The selection is based on secured \cone\ detection, 
availability of data at other wavelengths, and a \htwo\ surface 
density threshold as we explain below. The subset spans stellar 
masses in the range $10^{9.7}<\mstar/\msun<10^{11.5}$. Basic 
information is provided in Table~\ref{basic_table}.

\subsection{Molecular Gas and Star Formation Data}
We carry out our data analysis at two different resolutions: a fixed
angular resolution (6\arcsec), determined by our SFR indicator, and a
fixed spatial scale (1~kpc), to remove biases introduced by the
different distances to our objects. The high resolution near-infrared
and CO images were Gaussian-convolved to have the same angular as the
24~\um\ images. The 6\arcsec\ angular resolution covers a range of
physical scales in our sample, ${\cal R}\sim160-1250$~pc,
corresponding to physical distances $\rm D\sim5.5-43.1$~Mpc. The
galaxies NGC~772, NGC~3147, NGC~4273, and NGC~5371, shown in bold face
in Table \ref{basic_table}, have distances such that 6\arcsec\
corresponds to ${\cal R}\gtrsim1$~kpc in spatial scale --- those
galaxies are left at their native resolution in the 1~kpc analysis,
while the remaining 10 galaxies are Gaussian convolved to the angular
resolution corresponding to 1~kpc. All the analysis is carried out in
images that have been regrid to the same pixel-scale and sampled at 
Nyquist-rate.  We note here that, although we approximate the PSF of 
the MIR map as a Gaussian, the actual shape of the PSF is complex. It 
has prominent first and second Airy rings, with the second ring 
stretching out to $\sim$20\arcsec. Nevertheless, approximately 85\% 
of the total source flux is contained within the central peak with 
the FWHM of 6\arcsec\ (Engelbracht \etal\ 2007).

We construct molecular gas surface density (\Shtwo) maps and SFR
surface density (\Smir) maps in a similar manner as described in Paper
I. In the latter case following the prescription by Calzetti \etal\
(2007) who show that MIR 24~\um\ emission can be used as a SFR 
tracer for galaxies of normal metallicity and where the energy output 
is dominated by recent star formation. The MIR 24~\um\ SFR tracer is 
given by,  
\begin{equation}
\rm SFR (\msunyr) = 1.27 \times 10^{-38} \ \ [\nu L_{24}\ \ 
(\ergs) ]^{0.8850},
\label{sfrmir}
\end{equation}
where $\rm L_{24}$ (in $\rm erg\,s^{-1}\,Hz^{-1}$) is the luminosity
spectral density at 24~\um. We use a conversion factor $\xco=2.0
\times 10^{20}$~cm$^{-2}/{\rm (K km s^{-1})}$to determine
\htwo\ from \cone. The sensitivity ($1\sigma$) of the \Shtwo\ maps 
varies among galaxies from $\sim$1.0 to 5.1~\msunpc. The surface 
densities are multiplied by a factor of 1.36 to account for the 
mass contribution of helium.

\subsection{Stellar Masses and Surface Densities}
\label{stellarmass}

Construction of stellar surface density maps (\Sstar) involves two 
steps. First, we convert $\rm K_s$-band luminosity (\lstarks) to 
stellar mass (\mstar). We use the $\rm B-V$ optical colors for our 
galaxies and the relations by Bell \& de Jong (2003) 
to compute the mass-to-light ratios in each galaxy. In our sample, 
we find a mass-to-light ratio \mlrks$\sim0.3-0.8$~\msun/\lstarks.
Next, each pixel of the \lstarks\ map of a given galaxy is multiplied 
by the appropriate \mlrks\ to derive the stellar mass map, i.e., 
$\mstar=\mlrks \lstarks$.
Each pixel value of the \mstar\ map is then divided by the area of 
the pixel to obtain the \Sstar\ map. The galaxy integrated stellar
masses are determined by integrating their $H$-band 2-D surface
brightness profile after masking of foreground or background objects.
The NIR and MIR images were background subtracted prior to analysis. 
All surface densities have been inclination-corrected by applying a 
$\cos\,i$ factor.

There are uncertainties associated with the measurements of stellar
masses.  Although NIR emission is a good tracer of old (Gyr)
stellar populations and it experiences little internal extinction,
there is uncertainty associated with the determination of
mass-to-light ratios. For one, extinction affects the optical colors
used in the calibration and consequently the employed mass-to-light
ratio. More fundamentally, the mass-to-light ratio depends on the star
formation history of the disk. Hence it varies considerably among
galaxies of the same Hubble type and even within galaxies. For
example, young M-supergiants near the plane of the disk (Aoki \etal\
1991) as well as massive OB-associations (Regan \& Vogel 1994) can
contribute about $\sim50\%$ of the total NIR disk emission (Rix \& 
Rieke 1993; Regan \& Vogel 1994), potentially lowering the actual
mass-to-light ratio and causing us to overestimate masses by factors
of $\lesssim2$.

\subsection{Analysis}
The sensitivity and flux recovery of the interferometric map limits
the physical extent of the disk that can be studied. To minimize the
effects of deconvolution and flux recovery problems we analyze the
central 1\arcmin\ (in diameter) of the disks. Table \ref{basic_table} 
shows the physical extent of disk ($\mathcal S$) corresponding to the 
inner arcminute for the STING galaxies.  Following the methodology 
developed in Paper I we set a surface density threshold of 
$\Shtwo\gtrsim20$ \msunpc. 
Studying the emission from the bright regions of the molecular gas 
and SFR tracer maps ensures that: 1) the signal-to-noise is good,
2) interferometric deconvolution issues are minimized, 3) the
potential contribution by the diffuse emission (DE) is less
problematic, and 4) we focus on regions dominated by molecular gas. 
The DE is a component of the total disk emission that is unrelated 
to star formation activity and extended over the disk, in comparison 
to the localized emission associated with star formation. 
For example, a potential contributor to the DE at 24~\um\ is 
infrared cirrus emission. The \cone\ distribution of a galaxy can 
also contain DE not necessarily associated with the star-forming 
molecular clouds (Magnani \etal\ 1985; Blitz \& Stark 1986; Polk 
\etal\ 1988). Below our chosen surface density limit the DE in both 
SFR and molecular gas tracers has the potential to affect the 
molecular gas-SFR surface density relation (see Paper I for a 
detailed discussion on this issue).

Isolating the contribution of emission related to star formation
activity from widespread DE is a complex issue.
Recent work has shown that the DE has a potentially important impact
in the determination of the star formation law, depending on its
magnitude relative to emission coming from the star formation activity 
(Paper I; Liu \etal\
2011). In Paper I we used an unsharp masking technique to remove a
diffuse extended component, and showed that the most robust
measurements of the star formation law are those performed on the
bright regions of the studied galaxy (NGC~4254). In those regions the
contribution from DE is least significant, and the recovered star
formation law was approximately linear. Here we adopt the methodology
of Paper I, and minimize the impact of DE by focusing on the bright
regions of our sample of disks.  Finally, observational studies
suggest that \hone-to-\htwo\ phase transition occurs around
\Shtwo$\sim10-15$ \msunpc\ (Wong \& Blitz 2002; Bigiel \etal\ 2008;
Leroy \etal\ 2008) and the nature of total gas-SFR surface density
relation changes dramatically around this range (Kennicutt \etal\
2007; Bigiel \etal\ 2008; L08; Schruba \etal\ 2011 ). By focusing on
high molecular gas surface density regions we avoid this issue.

\begin{figure*}[t]
\epsscale{1}
\includegraphics{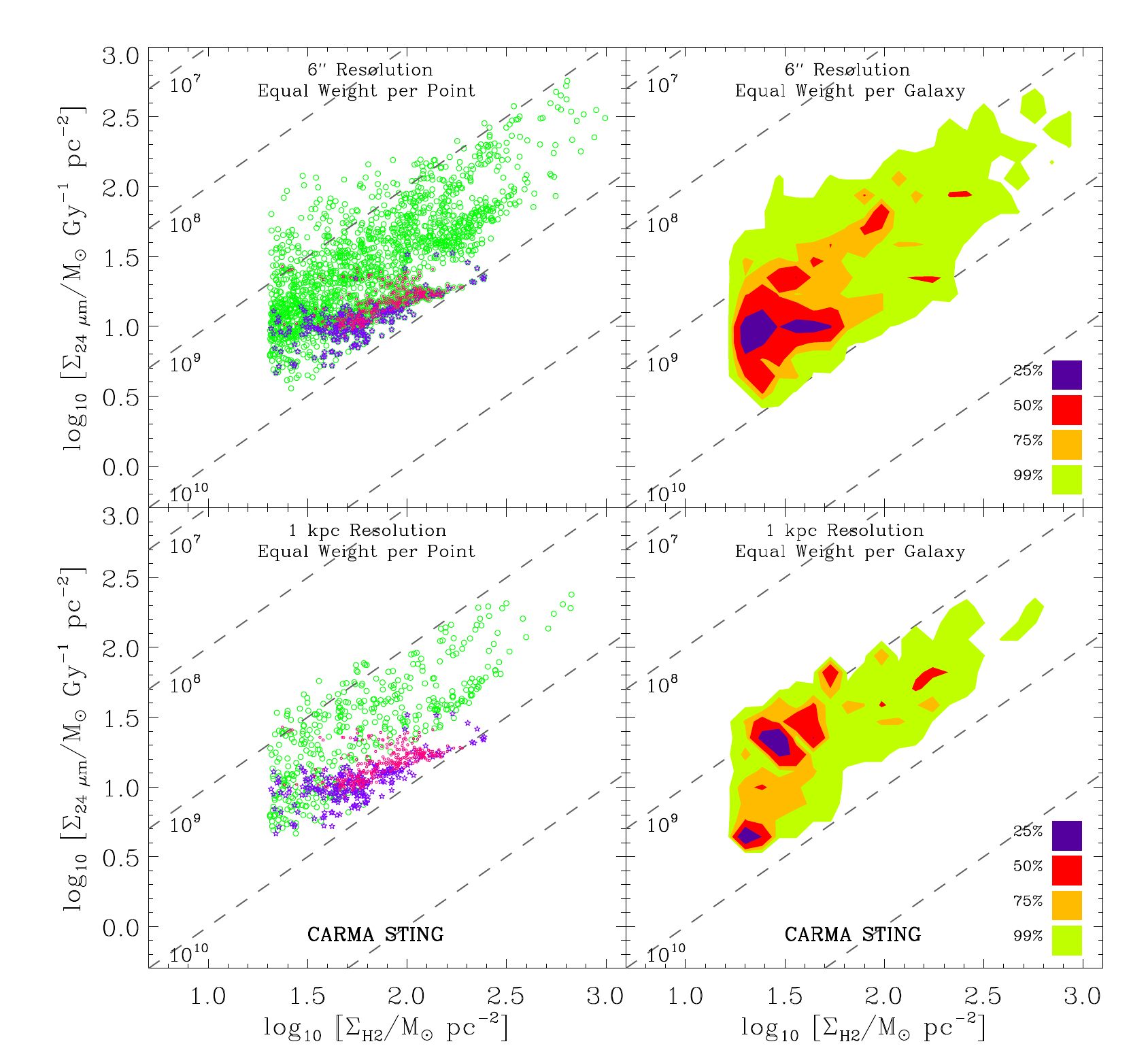}
\caption{Molecular gas-SFR surface density relation at 6\arcsec\ 
(top panels) and 1 kpc resolution (bottom panels) for the CARMA STING
galaxies. The left panels show scatter plots where each pixel has
equal weight in the distribution. The feature in the distribution
below $\log_{10}~[\Smir/\msunyrpc] \lesssim1.4$ showing a slow rise 
in \Smir\ with \Shtwo\ is due to NGC~772 (violet open star) and NGC~3147
(magenta open circle). The right panel shows a smoothed two-dimensional
distribution where each point is weighted by the inverse of the total
number of points of the contributing galaxy (all galaxies are equally
important in the distribution, irrespective of the number of points
they contribute). The contours of the smoothed distribution enclose
99\%, 75\%, 50\% and 25\% of the total. The diagonal dashed lines 
represent constant molecular gas depletion time \tdep, in years.
\label{ksplot}}
\end{figure*}

We carry out a pixel-by-pixel analysis (where the pixels are sampled 
at Nyquist-rate) to probe the highest possible spatial resolution. 
We use the Ordinary Least Square (OLS) bisector method (see Isobe 
\etal\ 1990) to fit the molecular gas-SFR surface density relation 
and account for the measurement errors in each variable. 
As discussed in Paper I, it is important to consider the impact of 
the sampling methodologies, fitting procedures, and measurement errors 
when determining the \Shtwo-\Ssfr\ relation. Some authors normalize 
\Shtwo\ at a characteristic surface density prior to regression analysis 
to minimize the covariance between the exponent and the normalization
constant of the power-law (see Bigiel \etal\ 2008 and Blanc \etal\
2009 in this regard). We do not apply such normalization, and we have 
verified that our results are robust to this choice.

\begin{figure*}[t]
\epsscale{1}
\includegraphics{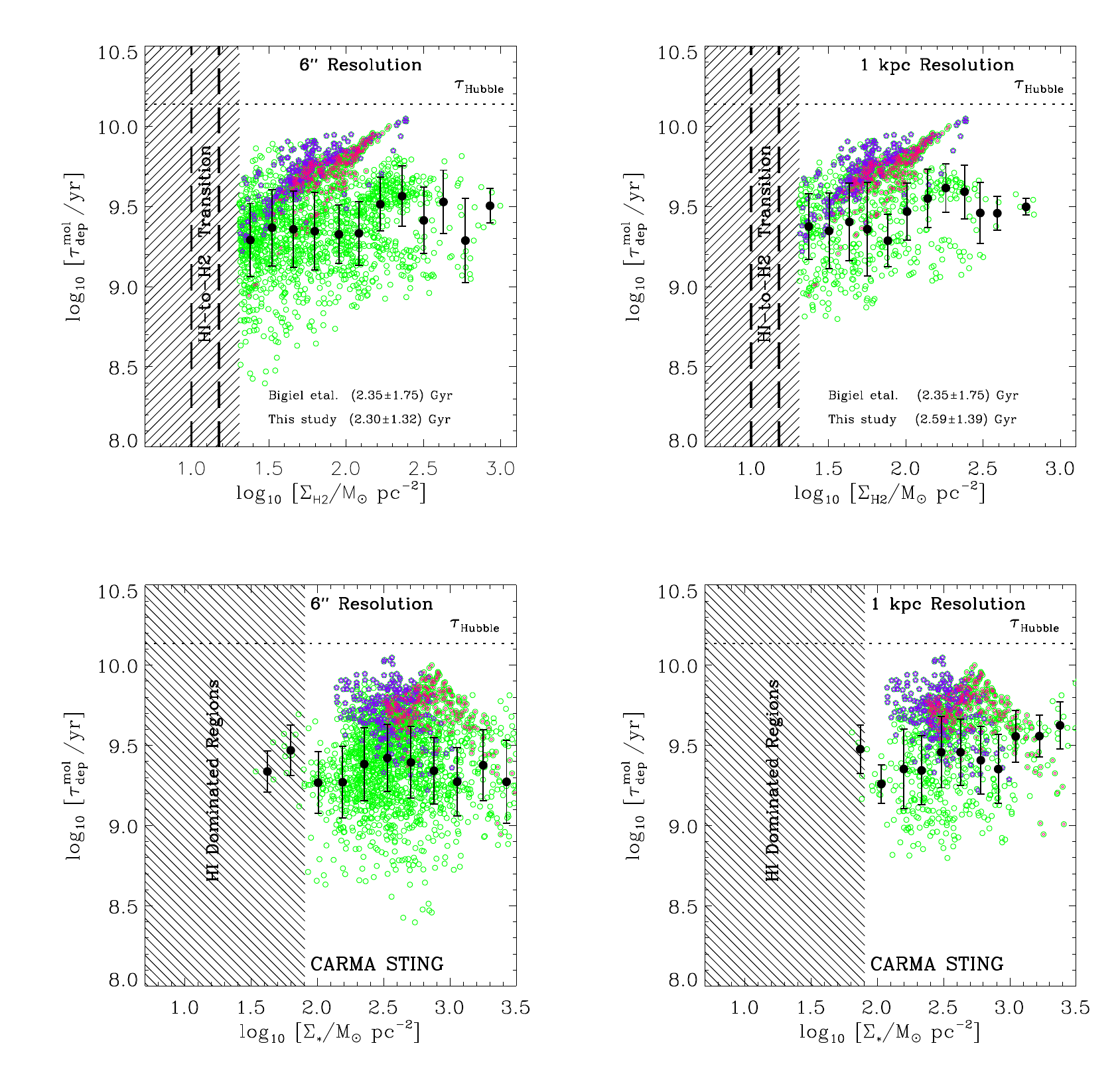}
\caption{Molecular gas depletion time (\tdep) versus molecular gas, 
\Shtwo, and stellar, \Sstar, surface density at 6\arcsec\ and 1~kpc 
resolution. The horizontal dotted line represents the Hubble time, 
and the filled circles and associated error bars in black represent 
the median and $1\sigma$ dispersion in \Shtwo\ (\Sstar) bins.  
{\em Top panels}: Results for \Shtwo. The gray hatch illustrates 
the region where $\Shtwo<20$~\msunpc, which we remove from the
analysis according to the discussion in \S\ref{analysis}. 
NGC~772 (violet open star) and NGC~3147 (magenta open circle) are 
responsible for the plume of points showing a slow rise in the 
depletion time with \Shtwo\ for $\Shtwo<50$~\msunpc. 
{\em Bottom panels}: Results for \Sstar. The gray hatch corresponds 
to the region where $\Sstar<80$~\msunpc, where \tdep\ increases for 
decreasing \Sstar\ in the L08 study. NGC~772 and NGC~3147 are 
colored similarly in these panels as in Fig. \ref{ksplot}. 
\label{dtplot}}
\end{figure*}

In this study we use the 24 \um\ emission as the SFR tracer because it
has several advantages over other single- or multi-wavelength tracers 
employed in the literature (Kennicutt \etal\ 2007; L08; Blanc \etal\ 
2009; Verley \etal\ 2010). Most importantly, 24 $\mu$m images of uniform 
quality are available for every galaxy in our sample, and no internal 
extinction correction is needed for this tracer. This is a major drawback 
for other SFR tracers at shorter wavelengths. Second, among various 
tracers studied in Paper I we find that the SFR obtained from 24~\um\ 
displays the tightest correlation with the molecular gas. The scatter 
varies in the range $\sim0.1-0.3$~dex depending on the subtraction of 
diffuse emission. Third, there is a striking spatial correspondence 
between the 24~\um\ and \cone\ maps (see also Rela\~no \& Kennicutt 
2009), suggesting that 24~\um\ is a faithful tracer of young (few 
million years), embedded star formation.

In the following discussion, we use \nmol\ to denote the power-law
index of the molecular gas star formation law, where the
\htwo\ gas includes the contribution from helium.

\section{Results}
\label{results}
\begin{figure*}[t]
\epsscale{1.2}
\includegraphics{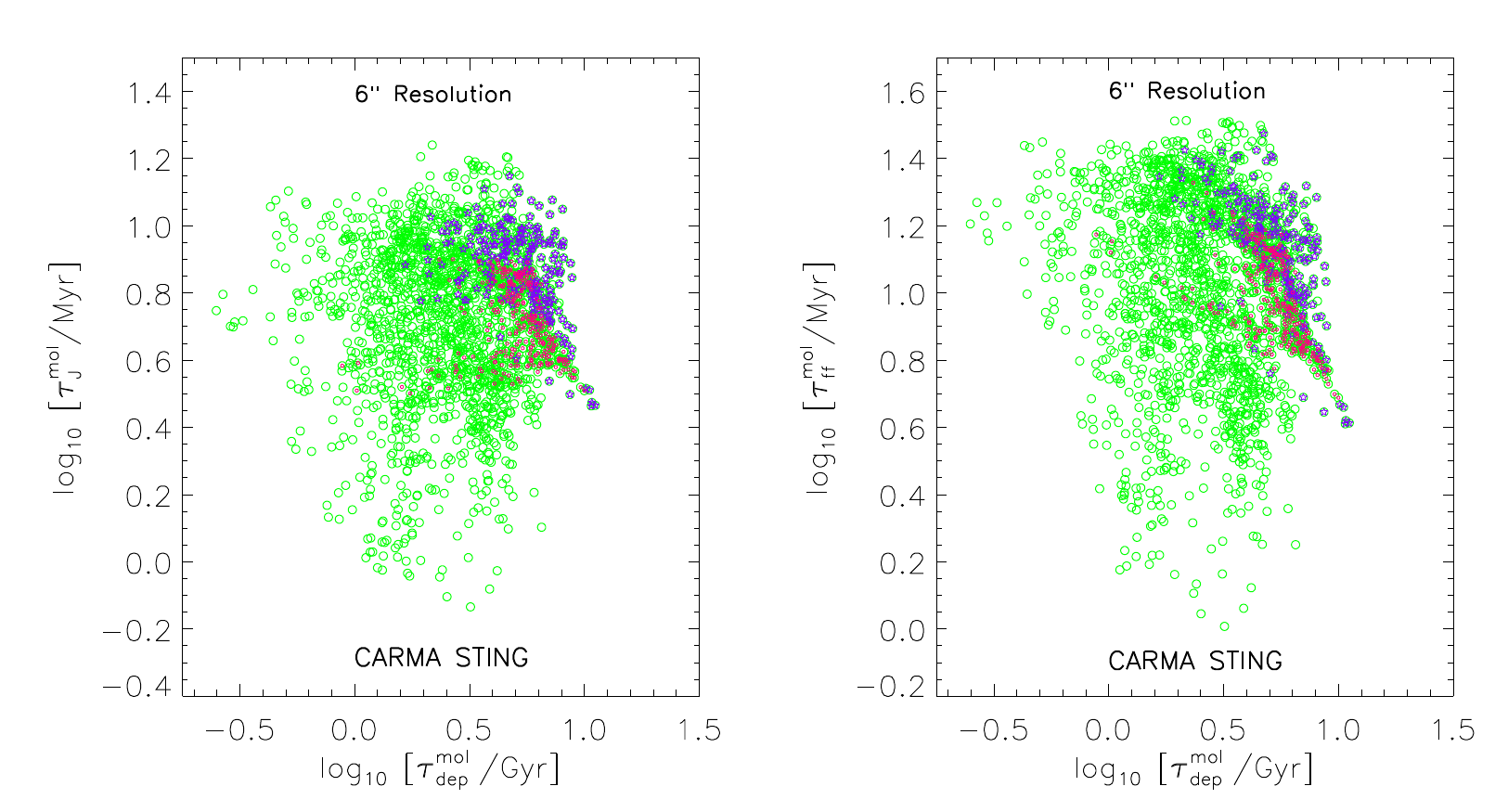}
\caption{Molecular gas depletion time (\tdep) versus local dynamical 
timescales.  {\em Left panel:} \tdep\ versus Jeans time (\tjeans).
{\em Right panel:} \tdep\ versus free-fall time (\tff). Results are
computed at 6\arcsec\ resolution. The measurements of NGC~772 and
NGC~3147 have similar color codes and legends as in Fig. \ref{ksplot}. 
There is
no correlation between \tdep\ and \tjeans, either on average or galaxy
by galaxy. By contrast, there is a reasonably strong correlation between 
\tdep\ and \tff\ on a galaxy-by-galaxy basis that is somewhat washed
out when considering the entire sample (see discussion in
\S\ref{dynamical}).
\label{dt_dt_plot}}
\end{figure*}

\begin{figure*}[t]
\epsscale{1}
\includegraphics{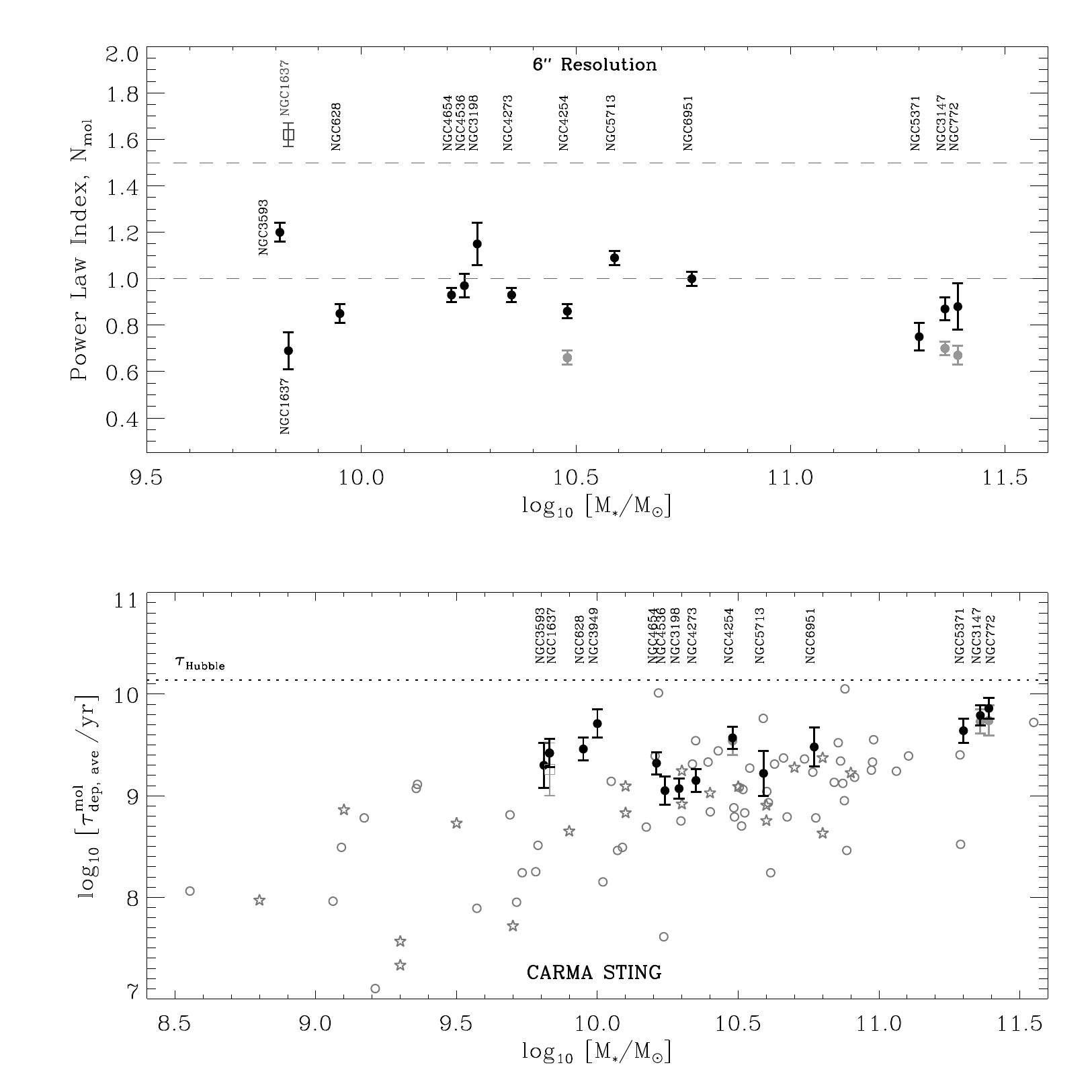}
\caption{The molecular star formation law as a function of galactic 
stellar mass. {\em Top panel}: Power-law index (\nmol) versus stellar
mass at 6\arcsec\ resolution. The black points and associated error
bars correspond to the determinations in our sample of galaxies. The
horizontal dotted lines represent indices for linear ($\nmol=1$) and
non-linear ($\nmol=1.5$) molecular gas star formation laws. The panel
also shows using a gray square \nmol\ for NGC~1637 if its nuclear
region is not masked. The measurements of NGC~772, NGC~3147,
and NGC~4254 shown in gray and black employ different \Shtwo\
thresholds (see \S\ref{global} for details).
{\em Bottom panel}: Molecular gas depletion time versus galactic 
(total) stellar mass. The average depletion time, \avetdep, is
calculated as the average of the logarithm of $\Shtwo/\Ssfr$ in
regions where \Shtwo\ is over the adopted threshold. 
The horizontal dotted line represents the Hubble time. The gray stars 
and circles show the global \tdep\ measurements within the optical 
disk of star forming galaxies taken from L08 and Kennicutt (1998), 
respectively. NGC~3949 has been shifted by +0.02 dex horizontally 
for clarity of presentation. \label{par_sm_plot}}
\end{figure*}

At 6\arcsec\ resolution our analysis includes all 14 galaxies. NGC~628
drops out of the sample at 1 kpc resolution, however, as its peak
surface density falls below the selected \Shtwo\ threshold after
smoothing. We have $\sim$2000 ($\sim$1000) Nyquist-sampled pixel
measurements arising from 14(13) galaxies at 6\arcsec\ (1 kpc)
resolution. The molecular gas and star surface densities span a wide
range, $\Shtwo\sim20 - 1000$~\msunpc\ and $\Ssfr\sim
4-570$~\msunyrpc. Our results are presented in Figures
\ref{ksplot}-\ref{par_sm_plot}.  The first three figures highlight
spatially resolved cases whereas Fig. \ref{par_sm_plot} shows global
quantities such as average depletion time (\avetdep, mean value of
\tdep\ within the central arcminute, calculated in the logarithm), 
and integrated stellar mass as given in Table~\ref{basic_table}.

\subsection{Molecular Gas Star Formation Law}

Figure \ref{ksplot} shows the molecular gas-SFR surface density
relation at 6\arcsec\ (top panels) and 1~kpc resolution (bottom
panels). The diagonal dashed lines represent constant molecular gas
depletion time defined as, $\tdep = \Shtwo/\Ssfr$, assuming zero 
recycling of the materials by massive stars into the ISM. The left 
panels show the scatter plots where each Nyquist-sampled pixel has 
equal weight in the distribution. NGC~772 (shown in violet) and 
NGC~3147 (magenta) create a feature in the distribution with a 
slower rise in \Smir\ with \Shtwo\ than other galaxies. We discuss 
this further in \S\ref{moldeptime}.

A simple correlation test shows that the points in this diagram are 
strongly correlated: at either resolution we find Spearman's rank 
correlation coefficient $\rho\sim0.7$. The OLS bisector method yields 
a power-law index \nmol~$\sim$1.1$\pm$0.1 at either resolution,
where the error is derived from bootstrapping.  Despite the fact these
measurements are coming from wide variety of galactic environments and
galaxy properties, the ensemble of points yields an approximately
linear star formation law. 

To right side panels of Figure \ref{ksplot} provide a view of the
sample that is not biased by galaxy size. In these panels any given
measurement is weighted by the inverse of the total number of
measurements of the galaxy to which the point belongs. Since larger
galaxies contribute more points to the ensemble, the weighting scheme
removes this bias by giving equal weight to each galaxy (Bigiel \etal\
2011). The contours enclose 99\%, 75\%, 50\% and 25\% of the
distribution. We note here that in extragalactic studies the inverse
of \tdep\ is sometimes known as the star formation efficiency (Young
\& Scoville 1991; McKee \& Ostriker 2007).

\begin{figure*}[t]
\epsscale{0.8}
\includegraphics{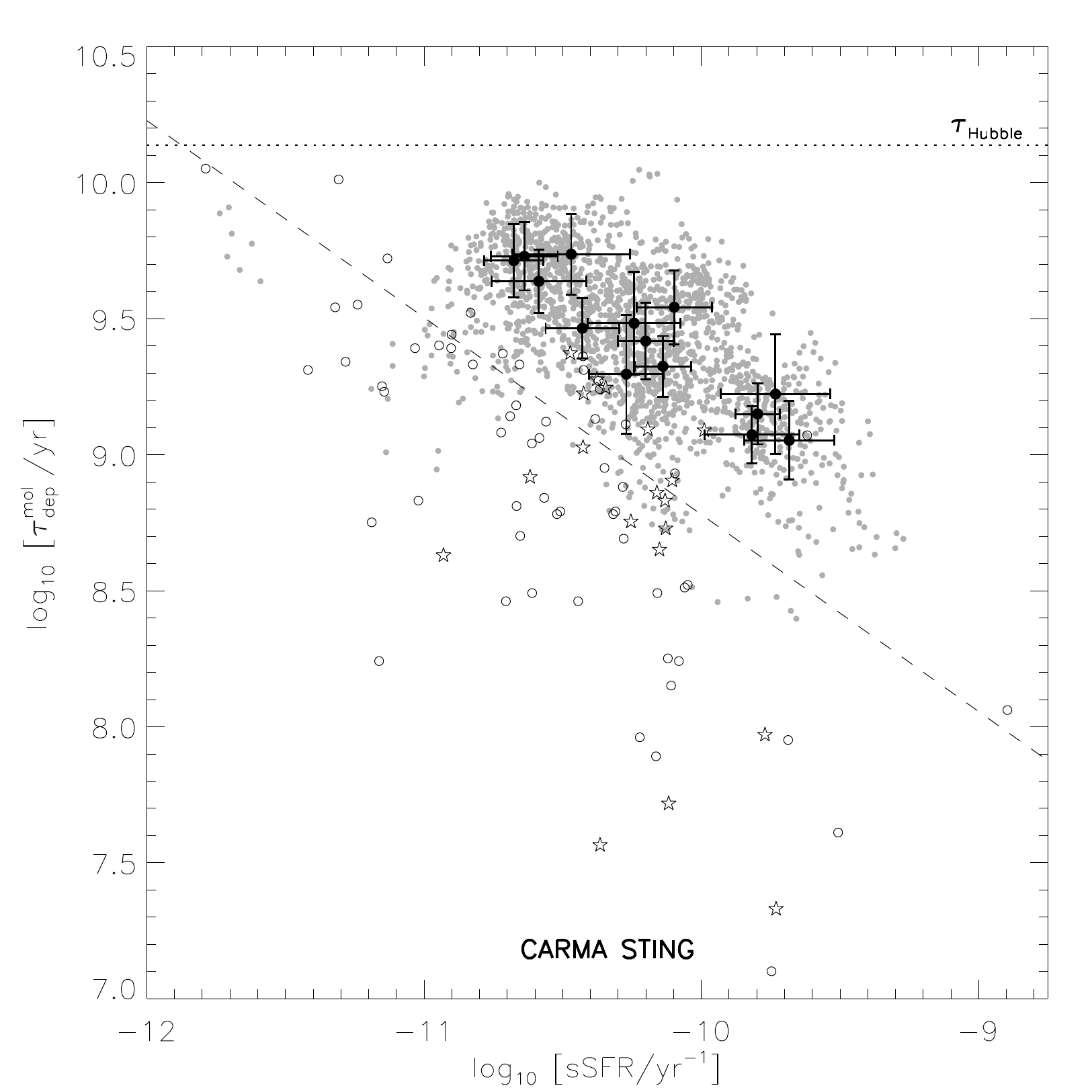}
\caption{Molecular gas depletion time versus specific star formation 
rate sSFR (in yr$^{-1}$) showing both resolved (6\arcsec) and global
measurements. The resolved measurements are shown in faint gray in the
background. The averages within the central arcmin and the associated
$1\sigma$ errors are shown with filled circles and thick black lines
respectively. The legends are similar to Fig. \ref{par_sm_plot}. The
horizontal dotted line represents the Hubble time and the diagonal
dashed line is the fit $\log[\tdep] = -0.724\,\log[\ssfr] + 1.54 $, 
from Saintonge \etal\ (2011). 
There are 7 outliers at the top-left corner associated with the 
spatially resolved measurements of NGC~5371. \label{dt_ssfr_plot}}
\end{figure*}

\subsection{Molecular Gas Depletion Time}
\label{moldeptime}
We show \tdep\ as a function of the molecular surface density \Shtwo\
in the top panels of Fig. \ref{dtplot}. The vertical hatch on the left 
of the panel demarcates the region where $\Shtwo<$20~\msunpc, the 
\Shtwo\ threshold discussed in \S\ref{analysis}. 
The individual measurements in STING galaxies are shown by green
points, with the binned medians and (1$\sigma$) dispersions in black
where the measurements from NGC~772 and NGC~3147 are excluded from the
bins (see below). A horizontal dotted line represents the Hubble
time, \thubble~$\sim$13.7 Gyr (Spergel \etal\ 2007). 

It is clear from these diagrams that the \htwo\ depletion time has at 
most very weak dependence on \Shtwo. A small correlation coefficient
($\rho\sim0.2$) suggests that \tdep\ is mostly uniform across the disk
The measurements from NGC~772 and NGC~3147 are responsible for the
plume of points showing a slow rise in the depletion time with \Shtwo\
for $\Shtwo<50$~\msunpc. Indeed the \Shtwo-\Ssfr\ relations in these
two galaxies show flatter than average power-law indices ($\nmol
\approx 0.8$).  The reason for this is unclear, but one possibility is
that the contribution of DE to their 24~\um\ luminosity is worse than
for the rest of the sample. Indeed \nmol\ is increasingly closer to 1
if we increase the \Shtwo\ threshold over 20~\msunpc. These galaxies
are two of the most distant, thus most massive and intrinsically
luminous, objects in our sample.  Excluding NGC~772 and NGC~3147 we
find $\tdep\sim2.30\pm1.32$~Gyr and $\tdep\sim2.59\pm1.39$~Gyr
at 6\arcsec\ and 1 kpc resolution, respectively (where the numbers
correspond to the median and $1\sigma$ dispersion in the logarithm of
the measurements). The depletion time increases slightly to
$\tdep\sim2.92\pm1.87$~Gyr and $\tdep\sim4.17\pm2.00$~Gyr at 
the respective resolutions when these two galaxies are included.
In any case with the adopted definition \tdep\ appears significantly 
shorter than the Hubble time.

Our measurements agree very well with the results by Bigiel \etal\ 
(2011), who studied the molecular gas depletion time in 30 nearby 
($<30$ Mpc) spirals from the HERACLES Survey (Leroy \etal\ 2009). 
Given the minimal overlap between the samples, the difference in the
criteria used to select the galaxies in both surveys, and the fact
that HERACLES is a single-dish \ctwo\ survey, this agreement shows 
that \tdep\ determined in the molecular gas is very similar in disks 
across a wide range of galaxy properties.

We show \tdep\ versus the stellar surface density \Sstar\ in the
bottom panels of Fig. \ref{dtplot}. The hatched section on the left
shows the region where \tdep\ decreases monotonically with increasing 
\Sstar, and \hone\ dominates the gas surface density (L08). 
We do not find any correlation between \tdep\ and \Sstar showing that 
the molecular gas consumption time is independent of stellar surface 
density in the region of the disk where \htwo\ is the dominant component 
of the ISM (see also L08 for a similar result). Since stellar mass 
surface density dominates the underlying gravitational potential in 
these objects, this result suggests that locally the molecular gas-SFR 
surface density relation is independent of the large scale galactic 
potential. In other words, once diffuse molecular clouds turns into 
isolated, self-gravitating objects, the conversion of \htwo\ into 
stars in GMCs is not sensitive to the overall gravitational potential 
GMCs reside.

Note that while both star formation rate and molecular gas mass surface 
densities correlate with \Sstar, their ratio \tdep\ does not correlate 
with either quantity. Both panels of Fig. \ref{dtplot} show \tdep\ is 
independent of \Shtwo\ and \Sstar. 
For the STING data set we find that the correlation coefficients for 
$\Smir - \Sstar$ and $\Shtwo - \Sstar$ relations are
$\rho\sim0.48$ and $\rho\sim0.63$, respectively, when all galaxies at
6\arcsec\ resolution are included. The correlation strengthens,
$\rho\sim0.65$ and $\rho\sim0.70$ for the respective relations, if 
the contributions from NGC~772 and NGC~3147 are removed from the
distribution of points. The correlation coefficients are similar at 
1~kpc resolution.

\subsection{Dynamical vs. Star Formation Timescale in the Molecular 
Disk} 
\label{dynamical}
Molecular gas and stellar mass surface densities as shown in Fig. 
\ref{dtplot}, along with the gas and stellar velocity dispersions, 
can be used to derive dynamical timescales associated with the growth
of GMCs such as the Jeans time (\tjeans) and the free-fall (\tff)
time. We would expect $\rm SFR \propto \Shtwo/\tjeans$ or 
$\Shtwo/\tff$ if either of these dynamical timescale is relevant
to star formation, which translates into a proportionality 
between \tdep\ and either dynamical time (L08; Wong 2009). 

For a plane parallel, axisymmetric, and isothermal two-component (gas 
and star) disk under hydrostatic equilibrium, the Jeans time can be 
written as,
\begin{equation}
\rm \tjeans = \frac{h_{z,g}}{c_g} 
= \frac{1}{\pi G} \left( \frac{c_g}{\Shtwo} \right) 
\left[ 1 + \frac{c_g}{c_\ast} \frac{\Sstar}{\Shtwo} \right]^{-1},
\end{equation}
under the assumption that the gas surface density is dominated by \htwo.
The quantities $\rm c_g$ and $\rm c_{\ast}$ are the velocity dispersion 
of gas and stars respectively along the $z$-direction, and $\rm h_{z,g}$ 
is the vertical scale height of the gaseous disk. The Jeans time defined 
this way should be regarded as the ``effective'' Jeans time since
originally it was defined in terms of thermal motions, with \cgas\
corresponding to the sound speed. Our \tjeans\ is defined in terms of
the gas velocity dispersion, which is dominated by turbulent motions.
The free-fall time of gas in the same disk can be written as,
\begin{eqnarray}
\rm \tff 
&=& \rm \sqrt{\frac{3\pi}{32} \frac{1}{G\rho_{H2}}} \nonumber \\
&=& \rm \frac{\sqrt{3}}{4G} \left( \frac{c_g}{\Shtwo} \right) 
\left[ 1 + \frac{c_g}{c_\ast} \frac{\Sstar}{\Shtwo} \right]^{-1/2},
\end{eqnarray} 
where $\rm \rho_{H2}$ is the mid-plane \htwo\ volume density,
related to the surface density by $\rm \rho_{H2} = \Shtwo/2h_{z,g}$.
These relations stem from the fact that in the two-component disk 
the scale height and the velocity dispersion of each component are 
interrelated (Kellman 1972; Talbot \& Arnett 1975; van der Kruit 
1983, 1988). For example, the relation between stellar velocity 
dispersion and scale height can be written as (see equation 16 in 
Talbot \& Arnett 1975),  
\begin{equation}
\rm c_\ast = \pi G \ h_{z,\ast} \ (\Sstar/c_\ast + \Shtwo/c_g).
\end{equation}

In order to estimate dynamical timescales we require both \cgas\ and 
\cstar, but direct measurement of either quantity is challenging.
We use the fact that \cstar\ and \hzstar\ are interrelated and so to 
obtain \cstar\ we first estimate stellar vertical scale height using 
the empirical relation between radial scale-length (\hrstar) and 
vertical scale-height (\hzstar) from Kregel \etal\ (2002), 
$\rm <h_{R,\ast}/h_{z,\ast}> = 7.3\pm2.2$, where \hrstar\ for STING 
galaxies is obtained from the 2MASS data. For \cgas\ we adopt a 
constant value of 10 \kmsec, very similar to that used by L08 and 
Wong (2009).

Figure \ref{dt_dt_plot} shows the correlations between \tdep, \tjeans, 
and \tff\ at 6\arcsec\ resolution. When the entire STING sample is 
considered we find no correlation between \tdep\ and either of the 
dynamical timescales (the correlation coefficients are $\rho\lesssim0.1$ 
for either timescale). Wong (2009) analyzed several star-forming 
galaxies employing radial profiles, and finding weak or no correlation 
between these quantities. Our results, with slightly different 
normalization for \tjeans\ and \tff\ and with a general solution of 
\cstar, corroborate these results in the molecular regions of disks 
for the ensemble of STING galaxies.

When considering individual galaxies, however, a clear negative 
correlation emerges between \tdep\ and \tff\ for most galaxies in the 
STING sample. Table 1 shows that this result is significant: the 
correlation coefficients for \tdep\ vs. \tjeans\ are small and 
scattered around 0 pretty much symmetrically, suggesting that the 
growth of Jeans-type dynamical instabilities is not responsible for the 
regulation of star formation. By contrast, for \tdep\ vs.\tff\ half the 
sample has $-0.7<\rho<-0.5$ and only one galaxy has $\rho\gtrsim0$. 
Although the sign of the correlation is puzzling, taken at face value 
this would suggests that some type of gravitationally driven accretion 
of molecular gas onto GMCs may be related to the regulation of star 
formation in the inner disks, and that the degree to which this plays 
a role may be different from galaxy to galaxy. But there are key 
caveats that make us skeptical of such conclusion.

We think that the observed correlation is likely not physical, but 
introduced by how we calculate \tff\ and \cstar, and the fact that 
moderate ($\rho\sim0.5$) to strong ($\rho>0.7$) correlations exist 
between \Shtwo, \Sstar, and 
\Ssfr. Thus there is a coupling between the X and Y axes computed in 
this manner that is difficult to avoid with the current state of the 
observations. To explore further we conduct a Monte Carlo experiment 
and use the fact that \Smir, \Shtwo, and \Sstar\ are interrelated 
(see the appendix for a detailed account of the numerical experiment). 
We approximately reproduce the observed correlations, \Sstar\ vs. 
\Smir\ and \Sstar\ vs. \Shtwo, assuming \Sstar\ as the independent 
variable. 
Our numerical experiment, although crude, strongly suggest that 
the observed galaxy by galaxy correlations may be almost entirely 
attributed to this effect. Additionally, there is considerable 
uncertainty in the degree of coupling between scale heights and 
velocity dispersions that complicates the interpretation.

\subsection{Global Correlations}
\label{global}
The top panel of Figure \ref{par_sm_plot} shows the distribution of
the molecular power-law index \nmol\ as a function of galaxy (total)
stellar mass \mstar. We only display the results at 6\arcsec\ 
resolution, as the results for a fixed 1~kpc spatial scale are 
effectively identical (see Table 1). Most of the galaxies show an 
approximately linear relation between molecular gas and SFR surface 
densities, with little indication of a trend with galaxy stellar 
mass. The STING sample average is $\nmol\sim0.96\pm0.16$. 
Note that the panel does not show the measurement of \nmol\ 
for NGC~3949, which has a very small dynamic range ($\sim$0.3 dex) 
in both \Shtwo\ and \Ssfr\ yielding a correlation coefficient close 
to zero.  

The largest value of the power-law index is found for NGC~1637,
$\nmol\sim1.6$, if we do not mask its nuclear region (gray square in
Fig. \ref{par_sm_plot}). As mentioned in section $\S$\ref{data},
despite not being a bona fide AGN the 24 $\mu$m image of NGC~1637
contains a point source at the center.  This source makes the
$\Shtwo-\Ssfr$ relation very steep when included. On the other hand,
when the central part of the is masked in order to remove the
contribution of the point source the power-law index becomes
significantly flatter ($\nmol\sim0.7$). This illustrates the potential
effect of AGN or nuclear clusters on the star formation law
determination, and some of the care that should be exercised to obtain
unbiased results (see also Momose \etal\ 2010).

We find significantly flatter than unity power-law indices
($\nmol\sim0.7-0.8$) in three other galaxies: NGC~772, NGC~3147, and
NGC~4254. The indices for these galaxies are shown in gray in the top
panel of Fig. \ref{par_sm_plot}. The reason for the low index in the
power-law fits is that in these sources \Smir\ is almost constant in
the range $\Shtwo\sim20-70$~\msunpc.  Consequently the distribution of
points in the $\Shtwo-\Ssfr$ plane possesses flat tails at the low
surface density end. Although the extended tail contains a small
number of points ($\lesssim20$ per galaxy), these points significantly
impact the regression analysis, yielding flatter indices. By contrast,
we obtain $\nmol\sim1$ in each case by raising the surface density
threshold to $\Shtwo\gtrsim70$~\msunpc, showing that the high surface
brightness regions display a linear trend.

The average molecular depletion time, \avetdep, in the STING galaxies
is shown in the lower panel of Fig. \ref{par_sm_plot} as a function of
galaxy stellar mass. The average is computed in the logarithm over the
pixels where \Shtwo\ is larger than the threshold.  For the STING
sample we find $\avetdep\sim3.19\pm1.91$~Gyr with a range of 1.1 to
7.2 Gyr for individual galaxies (see Table 1).  This sample average
\avetdep\ is slightly longer than the previous mentioned results for
the spatially resolved case, because of the ``per galaxy''
weighting. NGC~772 and NGC~3147 are again the contributing galaxies
that to push \avetdep\ to the larger value.

We observe a weak correlation between \avetdep\ and galaxy mass: a
linear regression analysis yields $\log \avetdep = (0.50\pm0.13)~\log
\mstar + (4.27\pm1.38)$. We should caution, however, that this
correlation is entirely driven by the high-mass galaxies ---
particularly NGC~772 and NGC~3147. 
The rank correlation coefficient between \avetdep\ and total \mstar\ 
is $\rho\sim$0.50 and $\sim$0.15 with and without these two galaxies. 
As discussed in the previous paragraphs, these are the galaxies 
where we also see a small \nmol. Note that although the derivation 
of stellar mass from NIR light has uncertainties (discussed in 
\S\ref{stellarmass}) the large dynamic range in \mstar\ lessens 
their impact on this study.

In a recent study, Saintonge \etal\ (2011) report an interesting 
correlation between \htwo\ depletion time and stellar mass in 
star-forming galaxies, in the stellar mass range 
$10^{10}~<\mstar/\msun<10^{11.5}$. 
Using unresolved molecular gas measurements and estimating SFRs 
using FUV+optical spectral energy distribution fitting, Saintonge 
\etal\ (2011) find a global disk-average \tdep~$\sim$1~Gyr in 
star-forming galaxies which increases by $\sim0.6$~dex for galaxies
with stellar masses from $\rm M_*\sim10^{10}$~\msun\ to $\rm
M_*\sim3\times 10^{11}$~\msun. By comparison the mean resolved
depletion time of the STING sample is $\avetdep\sim3.0\pm1.6$ Gyr,
consistent within the uncertainties with the depletion time found by
both L08 and Bigiel \etal\ (2011) who employ the same methodology. 

The difference between these two sets of results is due to the fact
that the studies measure fundamentally different quantities. The
resolved \tdep\ measurement is carried out in regions where CO is
detected, and it is thus a statement about the depletion time in GMCs
(particularly inner disk GMCs). The unresolved \tdep\ measurement
includes emission from all regions inside the beam, particularly the
outer disk, and it includes SFR tracer emission from regions that have
little or no molecular gas. It is thus a statement about the time it
would take for the galaxy to run out of molecular gas, modified by
variations in the \xco\ factor (which are likely to be significant for
outer disks) and contribution from diffuse emission and regions of
star formation that have no detectable CO emission to the global SFR
(which can be very significant in low surface brightness regions).

To explore further whether a $\avetdep-\mstar$ correlation is a
general property of star-forming disk galaxies we compile the global
measurements from L08 and Kennicutt (1998). This sample extends the
dynamic range in stellar mass by more than an order of magnitude
compared to Saintonge \etal\ (2011), incorporating many more low mass
systems $10^{8.5}~<\mstar/\msun\lesssim10^{11.5}$. The measurements 
of L08 and Kennicutt (1998) are shown by open stars and open circles,
respectively, in Fig. \ref{par_sm_plot}.
L08 use $\mlrks =0.5~\msun/\lstarks$ to derive the stellar mass for a
sample of 23 galaxies. This value is roughly in the middle of the
range of \mlrks\ used in this study. The SFR estimation and \xco\ used
by L08 are similar to ours, so no adjustment of the measurements are
necessary. The panel shows the measurements of 19 galaxies in L08 that
have secure \mhtwo\ estimates.
Kennicutt (1998) provides the global \Ssfr\ and \Shtwo\ of a sample 
of 61 star-forming galaxies. We adjust the original \Ssfr\ and \Shtwo\ 
measurements to be consistent those of L08 and this study. To derive 
\mstar\ for these galaxies we use $\mlrks =0.5~\msun/\lstarks$.
There is a minor overlap between the STING, and the L08 or Kennicutt
samples. On the other hand, there is a substantial overlap between 
the samples of L08 and Kennicutt. 
Since we do not intend to make any quantitative relation between
\avetdep\ and \mstar, we treat these samples as
independent because of their significant methodological differences.
It is apparent from Fig. \ref{par_sm_plot} that there is a trend in the 
global measurements, once the small mass galaxies are included. The global 
\tdep\ varies almost three orders of magnitude in the stellar mass range 
$10^{8.5}~<\mstar\lesssim10^{11.5}$, where small mass galaxies have 
systematically lower global \tdep. The trend for the small mass galaxies 
is simple to understand in terms of a higher value for \xco\ in galaxies 
that are of lower mass and lower metallicity (Leroy et al. 2011; Krumholz 
\etal\ 2011). A recent study of the Small Magellanic Cloud that avoids 
using CO to trace \htwo\ at low metallicity finds a strong metallicity 
effect in the CO emission, but not a measurable one in \tdep, suggesting 
that the trend in Fig. \ref{par_sm_plot} is mostly an \xco\ effect
(Bolatto et al. 2011).

Are systematic changes in \xco\ associated with metallicity strong enough 
to explain the magnitude of the observed trend? Reproducing the mean 
behavior of the unresolved data requires typically increasing \xco\ by 
a factor of 1.3 dex ($\sim$20) going from galaxies with stellar mass 
$\log \mstar \sim10.5$ to 9.0. The corresponding metallicity change 
expected, according to the mass-metallicity relation (Tremonti \etal\
2004; Mannucci \etal\ 2010) is approximately 0.6 dex. That is approximately 
the change in metallicity between the Milky Way (Baumgartner \& Mushotzky 
2006) and the Small Magellanic Cloud (Pagel 2003). Our best estimates of 
\xco\ on the large scales in the Small Magellanic Cloud show that it is 
30 to 80 times larger than Galactic (Leroy \etal\ 2011). If this behavior 
is typical, we conclude that \xco\ changes could indeed be the main driver 
behind the trend for global \tdep\ apparent in the lower panel of 
Fig. \ref{par_sm_plot}.

The past evolution of a galaxy is usually quantified by the the stellar 
mass assembly time (Kennicutt \etal\ 1994; Salim \etal\ 2007). It is 
the time required for a galaxy to assemble the current stellar mass at 
its present SFR.  Assuming zero recycling of the materials by massive 
stars into the ISM, this timescale is defined as, $\tsa = \Sstar/\Ssfr$. 
The inverse of this timescale is commonly known as the specific star 
formation rate (\ssfr).

Figure \ref{dt_ssfr_plot} shows \tdep\ versus \ssfr\ for the STING 
sample as well as the literature data discussed earlier. 
It also shows the relation between \tdep\ and \ssfr\ for star-forming 
galaxies in the local universe reported by Saintonge \etal\ (2011). 
Note that there are 7 outliers at the top-left corner associated with 
the spatially resolved measurements of NGC~5371 which are coming from 
its central regions where each of pixel has $\Sstar\gtrsim$2000~\msunpc.
Local galaxies form a bi-modal distribution in the \mstar~-~\ssfr\ 
plane (Salim \etal\ 2007) where star-forming galaxies form an 
horizontal branch. In terms of Fig. \ref{dt_ssfr_plot}, this observation, 
however, would imply that the large, massive galaxies would form a locus 
at the upper left whereas the low mass system would fall to the lower 
right, likely due the \xco\ effects discussed above. 
The global measurements from the literature are broadly consistent with 
the fit by Saintonge \etal\ (2011), with significant deviations at low 
masses. The resolved 6\arcsec\ (gray dots), 1~kpc (not plotted), and 
average (black symbols and error bars) measurements of STING galaxies 
show a similar correlation between \tdep\ and \ssfr\ ($\rho\gtrsim-0.6$). 
The  differences in the normalization between these measurements and 
the empirical relation reflects the difference in the aperture selection 
e.g., central arc-min vs. extended disk. 
As it can be surmised from the fact that \tdep\ is approximately constant 
with \Sstar\ (Fig. \ref{dtplot}), the negative correlation observed in the 
STING data set is likely attributable to the fact that the horizontal axis 
is proportional to the SFR whereas the vertical axis is proportional to 
the inverse of the SFR. We have further explored this correlations using 
a Monte Carlo experiment which suggests that the negative correlation 
between \tdep\ and \ssfr\ is governed at the fundamental level by the 
inter-connections among \Sstar, \Shtwo, and \Smir\ (see appendix for more 
on this experiment).   
\section{Discussion}
\label{discussion}
In this section we interpret the results of this study and relate them
to the existing scenarios of star formation and the molecular ISM. In
the spatially resolved case our results can be summarized as follows:
within the range of \Shtwo, \Sstar, and \mstar\ explored in this
study, 1) the resolved molecular gas depletion time is independent of
cloud properties in  the disk such as \Shtwo. 2) Dynamical timescales, 
such as the (effective) Jeans time and the free-fall time in the 
molecular disk, do not correlate with the molecular gas depletion time 
over the entire sample. And, 3) the resolved molecular gas depletion 
time is approximately independent of the disk environment represented, 
for example, by \Sstar\ or \mstar.

The uniformity of \tdep\ over a wide range of \Shtwo\ is most
naturally explained as the consequence of the approximate constancy of
the depletion time for molecular gas in GMCs.  Indeed, observations of
galaxies in the Local Group and beyond suggest that the properties
of GMCs are fairly uniform (Blitz \etal\ 2007; Bolatto \etal\ 2008; 
Fukui \& Kawamura 2010; Bigiel \etal\ 2010). In this scenario a linear 
star formation law follows naturally, where the $\Shtwo-\Ssfr$
relation arises from the number of GMCs filling the beam (Komugi
\etal\ 2005; Bigiel \etal\ 2008).  A linear molecular gas star
formation law is consistent with the scenario in which GMCs turn their
masses into stars at an approximately constant rate, irrespective of
their environmental parameters (Krumholz \& McKee 2005).

For our second result we determined two of the dynamical timescales
associated with the gravitational growth of GMCs in the molecular part
of the disk; the effective Jeans time and the free-fall time. This is
a challenging determination, because it relies on a number of
assumptions to estimate the relevant velocity dispersions and scale
heights.  The lack of a clear correlation between these dynamical
timescales and \tdep\ suggest that the star formation regulation is
not necessarily associated with the growth time of large scale
gravitational instabilities that may create and grow GMCs. We see a
correlation between \tdep\ and \tff\ in individual galaxies. Indeed,
approximately half the sample shows a correlation, although it is a
physically puzzling inverse correlation where longer free-fall times
correspond to shorter molecular depletion times, thus more star
formation activity per unit molecular mass. But our numerical
experiments suggest that where we see it, it is explained by the
observed correlations between \Shtwo, \Ssfr, and \Sstar.

The third result connects \tdep\ with the local gravitational
potential traced by \Sstar. This result implies that \tdep\ is mostly
independent of the local potential in the molecule rich regions of
the disk. L08 reaches a similar conclusion, and our result
corroborates their findings.  Following theoretical studies of
Elmegreen (1989, 1993) and Elmegreen
\& Padoan (1994) that rely on the equilibrium balance between \htwo\ 
formation and radiative dissociation, Wong \& Blitz (2002) and Blitz 
\& Rosolowsky (2004, 2006) suggested that the mid-plane hydrostatic 
pressure plays a critical role in governing the equilibrium fraction
of molecular gas phase in the disk. A recent theoretical study by 
Ostriker, McKee, \& Leroy (2010) explains the same observations as
due to the equilibrium in a multiphase ISM, where the stellar potential
plays also an important role. If the disk gravity is dominated
by the stellar potential and the gas scale height is smaller than the
stellar scale height, these studies show that the
molecular ratio, defined as $R_{\rm mol} = \Shtwo/\Shone$, increases
approximately linearly with the ambient pressure,

\begin{equation}
R_{\rm mol} \propto P^{\alpha} \propto (\Sgas \ \Sstar^{0.5})^{\alpha},
\end{equation}

\noindent 
where $\alpha\sim0.8-0.9$ and \Sgas\ is the mid-plane total gas 
(\hone+\htwo) surface density (L08). In the regions that are dominantly
molecular, however, $R_{\rm mol}\gg1$ and $\Shtwo\approx\Sgas$, thus
\tdep\ is independent of \Sstar. In other words, the stellar potential
plays a role at determining the fraction of gas that is molecular, but
it does not affect the rate at which GMCs collapse and their gas is
converted into stars. Blitz \& Rosolowsky (2004, 2006) show that the
ISM becomes molecular around $\Sstar\approx120\pm50$~\msunpc. The
mean (median) \Sstar\ of the STING galaxies is 580 (395)~\msunpc. It is,
therefore, very likely that we probe the molecular ISM where the
assumption $R_{\rm mol}\gg1$ is well justified (Xue \etal\ 2011, in 
preparation).

We should, however, stress here that whether the formation of molecular 
clouds is regulated by the mid-plane pressure (Blitz \& Rosolowsky 2004, 
2006), gravitational instability (e.g., Mac Low \& Glover 2010), or 
photo-dissociation (Krumholz et al. 2009) is still not settled. 
A combination of dynamical and thermodynamic factors may be required to 
regulate the formation GMCs and the subsequent star formation inside 
the clouds (e.g., Ostriker \etal\ 2010).

We find that the resolved molecular gas depletion time, averaged over
the central regions of our galaxies, shows a positive but weak
correlation with the integrated stellar mass in the stellar mass range
$10^{9.7}<\mstar/\msun<10^{11.5}$. Given that the correlation is
dominated by a few galaxies in this sample, it is difficult to assert
it with any degree of confidence. If it were real, it would suggest
that there are weak environmental effects on the \tdep\ in the GMCs,
and analytical models must be able to reproduce this large scale
behavior. A systematic resolved study with a large, well-defined
sample having homogeneous measurements over the extended disk is
necessary to shed more light on this issue.

Finally, using literature data we show that star forming galaxies,
spanning a large dynamic range in stellar mass
($10^{8.5}<\mstar/\msun\lesssim10^{11.5}$), show a clear correlation
between molecular depletion time and galaxy mass. This correlation is,
at least in the low-mass galaxies, most likely explained as arising
from systematic \xco\ variations, although systematic trends in the
SFR calibration may also play a role. Fundamentally, however, it
highlights different physical processes than the resolved molecular
gas measurements. The approximate constancy of \avetdep\ in the
resolved molecular measurements is a statement about the depletion
time in GMCs, which are regions of the disk dominated by molecular
gas. The unresolved \tdep\ measurement provides information about the
time it would take for the entire galaxy to run out of molecular gas 
at its current SFR, and it folds in the effects of the \xco\ factor 
and the contribution to the global SFR from emission arising in 
regions with no CO.

\section{Summary}
\label{summary}
We present a comprehensive analysis of the relation between \Shtwo\
and \Ssfr\ as a function of stellar mass at sub-kpc and kpc scales
using a sample of 14 nearby star-forming galaxies observed by the 
CARMA interferometer spanning the range of galaxy stellar masses
$10^{9.7}<\mstar/\msun<10^{11.5}$. We measure the relation in the bright,
high molecular gas surface density ($\gtrsim$20~\msunpc) regions of
star-forming disks. Sampling these CO-bright regions has the advantage of
minimizing the contribution from diffuse extended emission present in the 
SFR tracer and molecular gas disk. Our main results are:

\begin{enumerate}
\item The per-galaxy average star formation law for the sample,
determined using \Smir\ as the SFR indicator, is 
$\avetdep\sim3.19\pm1.91$~Gyr and $\nmol\sim0.96\pm0.16$.

\item The resolved molecular depletion
time \tdep\ is independent of both molecular and stellar surface
densities, \Shtwo\ and \Sstar, respectively. We find
$\tdep\sim2.30\pm1.32$~Gyr and $\tdep\sim2.59\pm1.39$~Gyr at 6\arcsec\
and 1~kpc resolution, respectively.

\item There is no clear correlation between \tdep\ and the effective
Jeans time, \tjeans, or the free-fall time, \tff, in the molecular
regions of our galaxies. These dynamical timescales, which may be
important for GMC growth, do not appear to regulate the star formation
once the gas is molecular.

\item There are no strong trends across our range of stellar masses for
either the power-law index or the normalization of the resolved
molecular star formation law.
\end{enumerate}

\acknowledgments
We thank the anonymous referee for useful comments and suggestions.
N. R. and A. B. acknowledge partial support from grants NSF
AST-0838178 and AST-0955836, as well as a Cottrell Scholar award from
the Research Corporation for Science Advancement. We thank the SINGS
team for making their outstanding data set available. This research
has made use of the NASA/IPAC Extragalactic Database (NED) which is
operated by the JPL/Caltech, under contract with NASA.
This publication makes use of data products from the 2MASS, 
which is a joint project of the University of Massachusetts and 
the IPAC/Caltech, funded by NASA and NSF.
Support for CARMA construction was derived from the Gordon and Betty 
Moore Foundation, the Eileen and Kenneth Norris Foundation, the Caltech 
Associates, the states of California, Illinois, and Maryland, and the 
NSF. Funding for ongoing CARMA development and operations are supported 
by NSF and CARMA partner universities. 
The National Radio Astronomy Observatory is a facility of the National 
Science Foundation operated under cooperative agreement by Associated 
Universities, Inc.
NGC~628, one of the STING galaxies studied in this paper, had been 
observed by CARMA as a part of the Ph.D. thesis of Misty La Vigne at 
the University of Maryland. 

\appendix
\section{Investigating Correlations Between Molecular Gas Depletion Time 
vs. Dynamical Timescales and Specific Star Formation Rate} 
\label{appen:montecarlo}


In this study we show that the observed trends in molecular gas 
depletion time vs. dynamical timescales for individual STING 
galaxies vary significantly, showing little or no correlation 
to strong negative correlation (see Table \ref{basic_table}). 
The correlations for either dynamical timescales, however, 
vanishes ($\rho\sim-0.01$ in \tdep~-~\tjeans\ and $\rho\sim-0.16$ 
in \tdep~-~\tff) when the entire ensemble is considered (see Fig. 
\ref{dt_dt_plot}). The molecular gas depletion time, on the other 
hand, shows strong negative correlation ($\rho\sim-0.64$) 
with the specific star formation rate (see Fig. \ref{dt_ssfr_plot}). 
In this appendix we demonstrate by a simple Monte Carlo experiment 
that these trends, particularly \tdep~-~\ssfr\ relation, are mostly 
driven by the existing correlations among \Shtwo, \Smir, and \Sstar\ 
and their connections to the stellar velocity dispersion. We carry 
out this experiment using the 6\arcsec\ resolution STING data set.  

\begin{figure*}
\epsscale{1}
\includegraphics{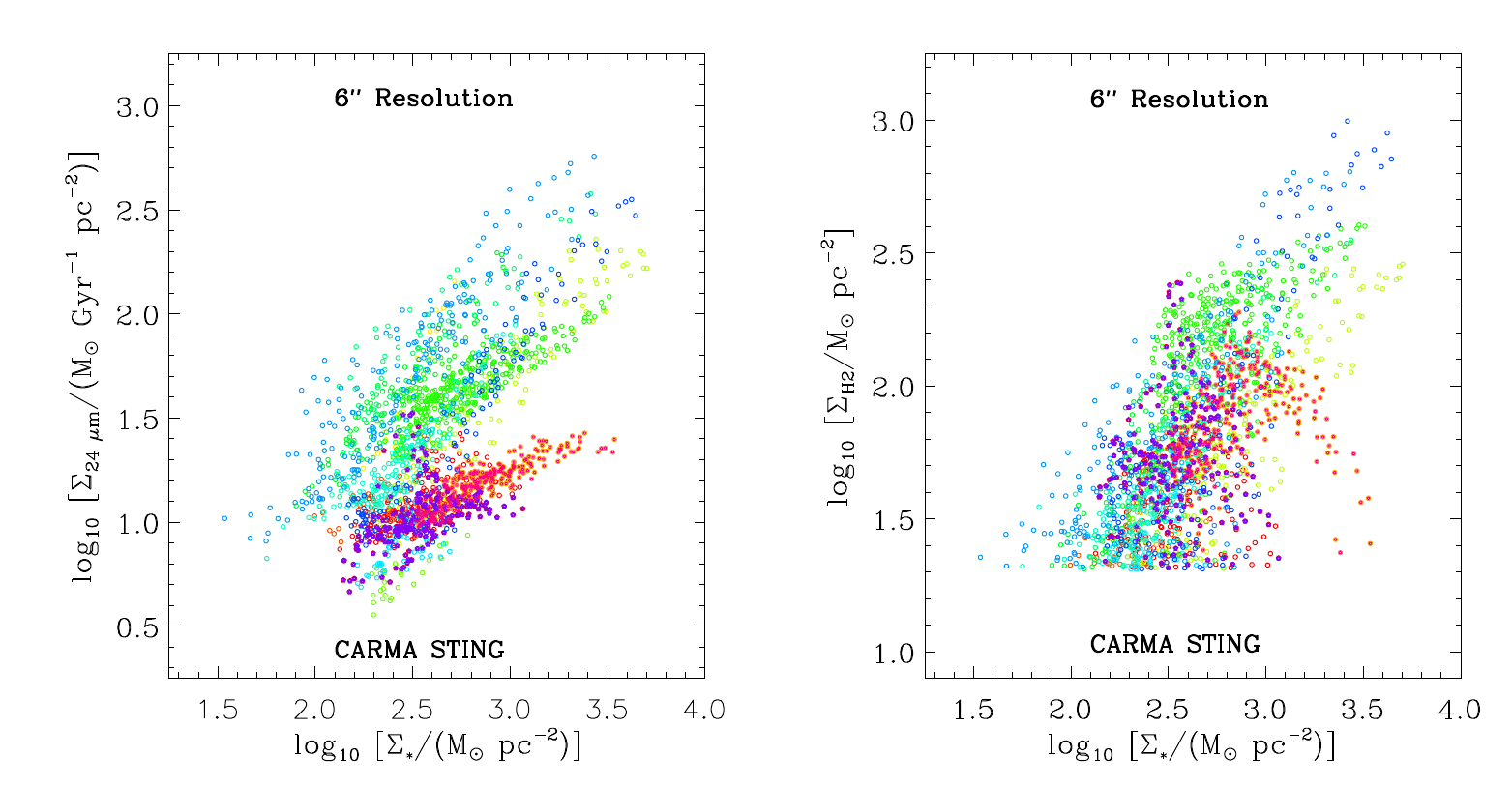}
\caption{The observed \Smir~-~\Sstar\ and \Shtwo~-~\Sstar\ correlations 
for STING galaxies. The measurements of each galaxy have different color 
code. The measurements of NGC~772 and NGC~3147 are highlighted using 
same color codes and legends as in Fig. \ref{ksplot}. 
\label{mc1_plot}}
\end{figure*}

We begin with the assumption that the stellar mass surface density 
is a fundamental parameter that influence both local and global 
evolution and organization of molecular gas in the disk and star 
formation. Indeed observational evidence shows that almost all 
physical variables strongly correlate with both global \mstar\ 
(Gavazzi 2009) and \Sstar\  (Dopita \& Ryder 1994) suggesting that 
both stellar mass and its surface density have significant roles in 
galaxy evolution. Strong positive correlations also exists between 
spatially resolved measurements of \Sstar, and \Shtwo\ and \Smir. 
Figure \ref{mc1_plot} shows such correlations ($\rho\sim0.5-0.8$) 
in both \Sstar~-~\Shtwo\ and \Sstar~-~\Smir\ diagrams for local 
measurements of STING galaxies. With the assumption about \Sstar\ 
made above, we express these relations by the following power laws,
\begin{equation} 
\rm \Ssfr = a \cdot \Sstar^m, 
\label{app:eqn1}
\end{equation} 
\begin{equation} 
\rm \Shtwo = b \cdot \Sstar^n,
\label{app:eqn2}
\end{equation} 
where $\rm (a,b)$ are normalization constants and $\rm (m,n)$ are 
power law indices. For each galaxy we derive these parameters using 
the OLS Bisector regression method in log space. We also obtain observed 
scatter (\sigobs) in each of these relations using the best-fit line.
These parameters are used as inputs in the Monte Carlo simulation and 
are shown in Table \ref{simulation_table}.

We use the transformation method to randomly sample stellar mass surface 
density from its observed distribution (see section 7.2 in Press \etal\ 
1992 for a detailed account of transformation method). This method 
provides three simple steps to generate random numbers from any arbitrary 
distribution function: 1) construction of the normalized 
cumulative distribution function of any given variable, 2) selection of 
a random number from a uniform distribution and inspection of the 
cumulative function to find the match between the two, and 3) selection 
of the variable within the domain that corresponds to the match in step 
2. We use this method with two additional conditions while sampling 
\Sstar\ from the observed distribution. These conditions are: 1) the 
lower and upper limits of the samples surface density must be within 
the range of observation, and 2) the number of observed measurements 
and simulated points must be the same. With a randomly sampled \Sstar, 
we generate SFR and molecular gas mass surface densities using, 
$\rm Y=\alpha \cdot X^{\beta} \cdot 10^{N(0,\sigobs)}$, where 
$\rm \alpha=\{a,b\}$, $\rm \beta=\{m,n\}$, and $\rm \{X,Y\}$ are the 
corresponding variables in equations \ref{app:eqn1} and \ref{app:eqn2}. 
The last term introduces scatter in the relation assuming a normally 
distributed \sigobs. Following this relation, we derive Y for any given 
X, which is then offset randomly using \sigobs. We repeat this 
for all the points contributed by any given galaxy.
For each galaxy we generate 5000 realizations and for each realization 
we compute stellar velocity dispersion, dynamical timescales, and \ssfr. 
A similar Monte Carlo method has been used previously in two-dimensional 
color-magnitude fitting analysis by Dolphin \etal\ (2001) and in 
constraining the \Ssfr~-~\Shtwo\ relation by Blanc \etal\ (2009). 

NGC~3949 has a very small dynamic range ($\sim$0.3 dex) in both \Shtwo\ 
and \Smir\ yielding a negative correlation in the \Shtwo~-~\Sstar\ plane 
and a correlation coefficient close to zero in the \Shtwo~-~\Smir\ plane. 
We exclude this galaxy from our simulation. It contributes only 2\% to 
the ensemble of points and has virtually no impact on our results. 
We also exclude the seven outliers with $\Sstar\gtrsim2000$~\msunpc\ 
from NGC~5371 data.

\begin{figure}
\epsscale{0.80}
\includegraphics{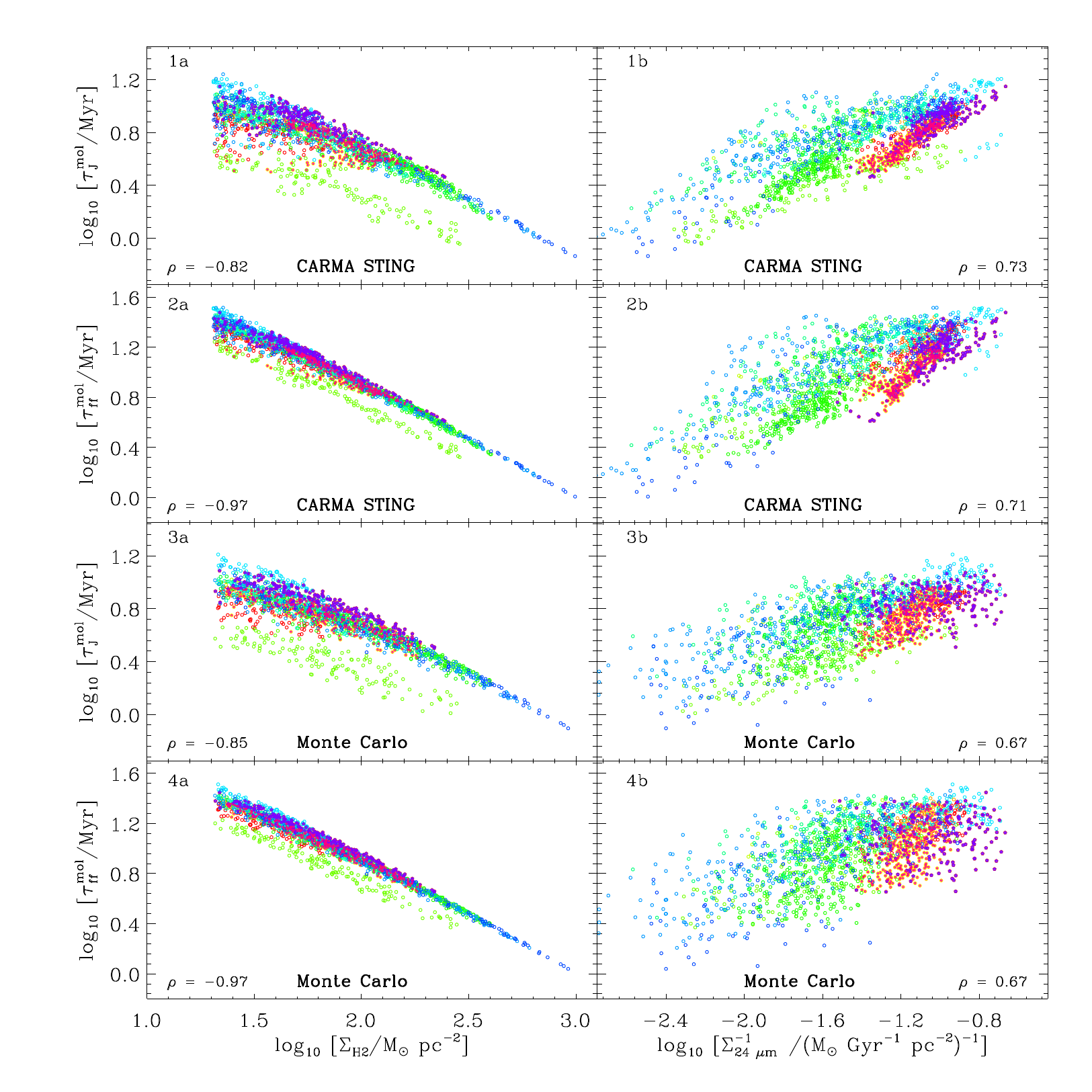}
\caption{Observations vs. simulations for STING galaxies. 
The dynamical timescales as a function of \Shtwo, and inverse \Smir\ are 
shown in panels 2 and 3. These timescales derived from a single realization 
of the Monte Carlo simulation are shown in panels 4 and 5. Each panel shows 
the Spearman rank correlation coefficient ($\rho$) for the corresponding 
relationship. 
The observed and simulated measurements of each galaxy have the same color 
code as in Fig. \ref{mc1_plot}. The measurements of NGC~772 and NGC~3147 
are highlighted using same color codes and legends as in Fig. \ref{ksplot}. 
\label{mc2_plot}}
\end{figure}

\begin{figure*}
\epsscale{0.80}
\includegraphics{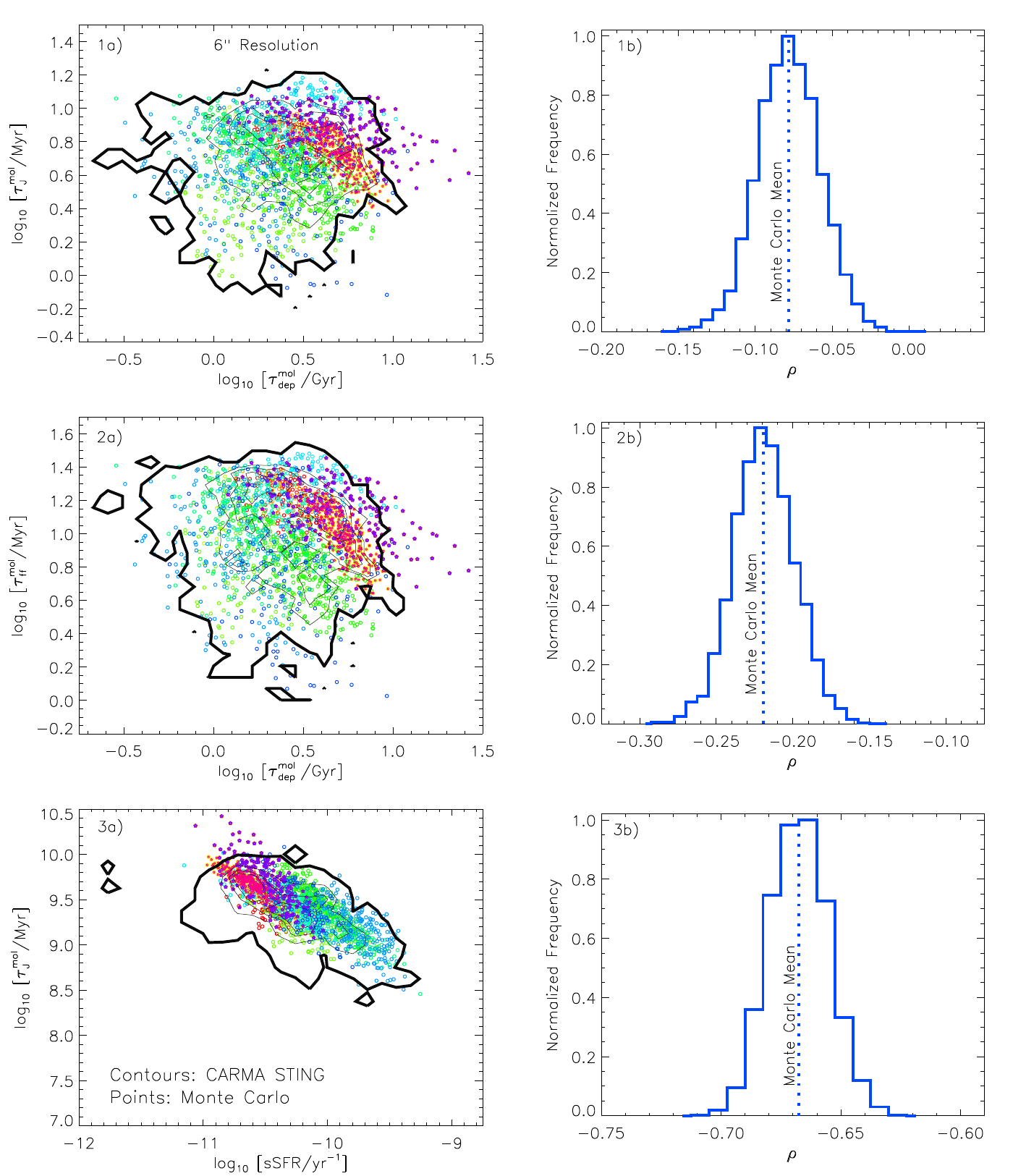}
\caption{Observations vs. simulations of STING galaxies. The \tdep~-~\tjeans, 
\tdep~-~\tff, and \tdep~-~\ssfr\ relations for observed and simulated 
measurements are shown, respectively, in panels (1a), (2a), and (3a) on the 
left. The open contours (inwards) enclose 99\%, 75\%, 50\% and 25\% of the 
observed distribution. The simulated measurements comes from the same 
realization and have the same color code as in Fig. \ref{mc2_plot}. 
The right panels show the normalized distribution functions of the Spearman 
correlation coefficient obtained from 5000 random realizations of the data. 
The mean ($\bar{\rho}$) and the standard deviation ($\sigma_{\rho}$) of the 
corresponding distribution functions are -0.08$\pm$0.02, -0.22$\pm$0.02, 
and -0.68$\pm$0.01. The correlation coefficient for the corresponding 
observed relations are -0.01, -0.16, and -0.64. 
\label{mc3_plot}}
\end{figure*}

The results of our simulations are presented in Figs. \ref{mc2_plot} 
and \ref{mc3_plot}. Figure \ref{mc2_plot} highlights various observed 
relations and their reproductions from a single realization. While 
panels 1 and 2 demonstrates relationships between dynamical timescales 
and \Shtwo\ and \Smir\ for the observed data, panels 3 and 4 show these 
timescales for simulated measurements. Simple correlation analysis 
strongly suggests that the simulation reproduce the observation reasonably 
well. 
Simulated measurements, however, show slightly larger scatter ($\sim$0.15 
dex in panels 1b and 3b as compared to $\sim$0.25 dex in panels 2b and 4b), 
which is mostly due to \sigobs. Since \sigobs\ is a quadratic sum of 
intrinsic scatter (\sigint) and measurement error (\sigmes), in absence 
of \sigmes\ the observed scatter would always overestimate the true scatter 
present in any relation.

Figure \ref{mc3_plot} shows all three major correlations for both observed 
and simulated measurements. For the purpose of demonstration the simulated 
data are drawn from the same realization as in Fig. \ref{mc2_plot}. 
A comparison of the correlation coefficients of individual galaxies in 
Tables \ref{basic_table} and \ref{simulation_table} shows that the simulation 
approximately reproduces the observed \tdep~-~dynamical timescale correlations. 
The simulation also produces \tdep~-~\ssfr\ correlation comparable 
($-0.71<\rho<-0.62$) to the observed one ($\rho\sim-0.64$).   
The right panels show the normalized distribution functions of the correlation 
coefficient derived from 5000 realizations. 
The mean ($\bar{\rho}$) and the standard deviation ($\sigma_{\rho}$) of the 
corresponding distribution functions are (-0.08$\pm$0.02), (-0.22$\pm$0.02), 
and (-0.68$\pm$0.01). 
The mean values of the respective distribution functions reflect the intrinsic 
strengths of these correlations. For example, a small correlation coefficient 
($\rho\lesssim0.2$) derived from a large set of measurements indicates no 
correlation between \tdep\ with either Jeans timescale or free-fall timescale. 
Likewise, one can find a strong anti-correlation between \tdep and \ssfr. 

The properties of the distribution functions shown in Fig. \ref{mc3_plot} 
depends on how scatter is introduced in Eqs. \ref{app:eqn1} and \ref{app:eqn2}. 
While simulation of each individual galaxy takes normally distributed \sigobs\ 
as an input, the observed scatter, however, is not symmetric and varies with 
surface density (see Fig. \ref{mc1_plot}). An outcome of this simple choice of 
\sigobs\ is the narrow distribution functions of $\rho$ and the offsets between 
$\bar{\rho}$ and observation. However, even with this simplistic approach we 
approximately reproduce the observed strengths of various correlations using 
Eqs. \ref{app:eqn1} and \ref{app:eqn2} which suggests that these relations are 
mostly determined by the interrelations among \Smir, \Shtwo, and \Sstar\ and 
\cstar.


\begin{deluxetable*}{@{}lccccccccc}
\tablecaption{Input and Output Parameters of Monte Carlo Simulations}
\tabletypesize{\scriptsize}
\tablewidth{7.0truein}
\tablecolumns{10}
\tablehead{
\colhead{Object}  
&\colhead{$\rm \log a$}    
&\colhead{$\rm m$}
&\colhead{\sigobs}
&\colhead{$\rm \log b$}    
&\colhead{$\rm n$}
&\colhead{\sigobs}
&\colhead{N} 
&\colhead{Corr.} 
&\colhead{Corr.} \\
\colhead{}
&\colhead{} 
&\colhead{}  
&\colhead{dex}
&\colhead{}
&\colhead{}
&\colhead{dex}
&\colhead{}
&\colhead{$\rho$}
&\colhead{$\rho$} \\
 \colhead{(1)}
&\colhead{(2)} 
&\colhead{(3)}
&\colhead{(4)}  
&\colhead{(5)}  
&\colhead{(6)}
&\colhead{(7)}
&\colhead{(8)}
&\colhead{(9)}
&\colhead{(10)}
} 
\startdata 
NGC~~628    &-0.50     &0.62     &0.11     &-0.81     &~0.92    &0.11     &131    &-0.23    &-0.45\\
NGC~~772    &-1.12     &0.85     &0.19     &~0.16     &~0.63    &0.19     &217    &-0.40    &-0.54\\
NGC~1637    &-0.80     &0.82     &0.09     &-1.45     &~1.27    &0.09     &~47    &-0.72    &-0.76\\
NGC~3147    &-0.20     &0.49     &0.06     &-0.65     &~0.90    &0.06     &298    &-0.79    &-0.82\\
NGC~3198    &-3.65     &2.13     &0.08     &-2.99     &~1.89    &0.08     &~18    &0.41     &+0.24\\
NGC~3593    &-1.53     &1.08     &0.17     &-0.74     &~0.89    &0.17     &141    &0.19     &-0.04\\
NGC~3949    &\ldots    &\ldots   &\ldots   &\ldots    &\ldots   &\ldots   &~27    &\ldots   &\ldots\\
NGC~4254    &-0.14     &0.64     &0.10     &-0.58     &~0.99    &0.10     &308    &-0.64    &-0.67\\
NGC~4273    &-1.10     &1.12     &0.09     &-1.19     &1.22     &0.09     &103    &-0.10    &-0.16\\
NGC~4536    &-0.44     &0.89     &0.15     &-0.52     &0.94     &0.15     &~67    &-0.16    &-0.28\\
NGC~4654    &-1.05     &0.96     &0.10     &-0.93     &1.03     &0.10     &168    &-0.20    &-0.32\\
NGC~5371    &-1.41     &0.94     &0.11     &-1.50     &1.22     &0.11     &~58    &-0.57    &-0.61\\
NGC~5713    &-0.94     &1.07     &0.21     &-0.59     &0.99     &0.21     &220    &-0.12    &-0.20\\
NGC~6951    &-1.98     &1.26     &0.18     &-1.54     &1.26     &0.18     &135    &-0.10    &-0.17        
\enddata 
\tablecomments{Column $(1)$: Galaxy name. 
Column $(2)$, $(3)$ \& $(4)$: Normalization constant, power-law index, and 
observed scatter (in dex) derived from \Smir~-~\Sstar\ relation. 
Column $(5)$, $(6)$ \& $(7)$: Normalization constant, power-law index, and 
observed scatter (in dex) derived from \Shtwo~-~\Sstar\ relation.  
Column $(8)$: Number of independent observed measurements.  
Column $(9)$ \& $(10)$: Spearman rank correlation coefficients ($\rho$) from 
\tdep~-~\tjeans\ and \tdep-\tff\ relations, respectively, for one realization 
of Monte Carlo simulation.  \label{simulation_table}}
\end{deluxetable*}

\end{document}